\documentclass[sigconf]{acmart}

\AtBeginDocument{%
  }
    
\copyrightyear{2025}
\acmYear{2025}
\setcopyright{cc}
\setcctype{by-nd}
\acmConference[CHI '25]{CHI Conference on Human Factors in Computing Systems}{April 26-May 1, 2025}{Yokohama, Japan}
\acmBooktitle{CHI Conference on Human Factors in Computing Systems (CHI '25), April 26-May 1, 2025, Yokohama, Japan}\acmDOI{10.1145/3706598.3714230}
\acmISBN{979-8-4007-1394-1/25/04}

\newcommand{\change}[1]{{\color{black} #1}}
\newcommand{\cameraready}[1]{{\color{black} #1}}

\newcommand{\remove}[1]{}

\graphicspath{{figures/}}

\usepackage{longtable}
\usepackage{graphicx}
\usepackage{subcaption}
\usepackage{multirow}
\usepackage{array}
\usepackage{booktabs}
\usepackage{geometry}
\usepackage{makecell}
\usepackage{soul}
\usepackage{float}

\usepackage{afterpage}

\begin{document}


\title[Inclusive Avatar Guidelines]{Inclusive Avatar Guidelines for People with Disabilities: Supporting Disability Representation in Social Virtual Reality}


\author{Kexin Zhang}
\affiliation{
    \institution{University of Wisconsin-Madison}
    \city{Madison}
    \state{Wisconsin}
    \country{USA}
}
\email{kzhang284@wisc.edu}

\author{Edward Glenn Scott Spencer}
\affiliation{
    \institution{Virginia Tech}
    \city{Blacksburg}
    \state{Virginia}
    \country{USA}
}
\email{scottspencer@vt.edu}

\author{Abijith Manikandan}
\affiliation{
    \institution{Virginia Tech}
    \city{Blacksburg}
    \state{Virginia}
    \country{USA}
}
\email{abijith@vt.edu}

\author{Andric Li}
\affiliation{
    \institution{University of California San Diego}
    \city{La Jolla}
    \state{California}
    \country{USA}
}
\email{arl009@ucsd.edu}

\author{Ang Li}
\affiliation{
    \institution{University of Wisconsin-Madison}
    \city{Madison}
    \state{Wisconsin}
    \country{USA}
}
\email{ali253@wisc.edu}

\author{Yaxing Yao}
\affiliation{
    \institution{Virginia Tech}
    \city{Blacksburg}
    \state{Virginia}
    \country{USA}
}
\email{yaxing@vt.edu}

\author{Yuhang Zhao}
\affiliation{
    \institution{University of Wisconsin-Madison}
    \city{Madison}
    \state{Wisconsin}
    \country{USA}
}
\email{yzhao469@wisc.edu}

\renewcommand{\shortauthors}{Zhang et al.}

\begin{abstract}
Avatar is a critical medium for identity representation in social virtual reality (VR). However, options for disability expression are highly limited on current avatar interfaces. Improperly designed disability features may even perpetuate misconceptions about people with disabilities (PWD). As more PWD use social VR, there is an emerging need for comprehensive design standards that guide developers and designers to create inclusive avatars. Our work aim to advance the avatar design practices by delivering a set of centralized, comprehensive, and validated design guidelines that are easy to adopt, disseminate, and update. Through a systematic literature review and interview with 60 participants with various disabilities, we derived 20 initial design guidelines that cover diverse disability expression methods through five aspects, including avatar appearance, body dynamics, assistive technology design, peripherals around avatars, and customization control. We further evaluated the guidelines via a heuristic evaluation study with 10 VR practitioners, validating the guideline coverage, applicability, and actionability. Our evaluation resulted in a final set of 17 design guidelines with recommendation levels.

\end{abstract}




\maketitle

\section{Introduction}

Social Virtual Reality (VR) has gained increasing popularity in the online social ecosystem, where multiple users can meet, interact, and socialize in the form of avatars \cite{Freeman2020}. Unlike 2D social media, social VR is equipped with body-tracking avatars, synchronous voice chat, and kinesthetic interactions. These features enabled rich communication modalities through the avatar's body gestures, facial expressions, and even gaze interactions \cite{maloney_nonverbal}. Many users thus view the avatar as an extension of themselves and a major medium for self-expression \cite{Manninen2007, Freeman2021}.

People with disabilities (PWD), a prevalent yet marginalized community \cite{WHO2023}, have shown increasing presence in social VR \cite{zhang2022, meta_horizon_accessibility}. 
Like any other user group, PWD curate social images by customizing avatars, and many prefer disclosing their disabilities via avatars as a core part of their identities \cite{zhang2022, kelly2023}. Previous research has identified PWD's needs and strategies to express disabilities in virtual worlds and social VR spaces \cite{zhang2022, kelly2023, Gualano_2023, zhang2023}. For example, Zhang et al. \cite{zhang2022} interviewed PWD and revealed that they preferred to disclose disabilities through various strategies, such as showing disability authentically or presenting a certain aspect of their disability selectively. Some other works \cite{kelly2023, Gualano_2023} focused on people with invisible disabilities (e.g., ADHD) and identified their preferences and needs for disability expression in various social VR contexts. 

Despite the tremendous needs, enabling PWD to flexibly and properly express their disabilities remains challenging \change{in practice}. \change{Many social VR platforms do not support disability representation ~\cite{zhang2022}}, and some avatar designs for PWD can be misleading (e.g., avatar on a virtual hospital wheelchair \cite{kelly_AI24, zhang2023}), further perpetuating misconception and bias towards disability. \change{While prior research has derived various design recommendations for disability representation \cite{zhang2022, kelly2023, assets_24}, they were dispersed across multiple venues, making it challenging for practitioners to access and apply comprehensively}. \change{Moreover, no research has systematically validated these design recommendations.} 
\change{There is a pressing need for a centralized, comprehensive, and validated set of design guidelines that are easy to adopt, disseminate, and update \cite{human_ai_guidelines}, facilitating the practical integration of inclusive avatar design into social VR development.}

To fill this gap, we generated and evaluated a comprehensive set of avatar guidelines to provide actionable guidance and examples for VR developers and designers, enabling them to integrate avatar features for suitable disability representation without bringing bias and misconceptions. To achieve this goal, we conducted an interview with 60 PWD with diverse types of disabilities to deeply understand their preferred disability expression methods and the potentially biased designs they want to avoid. \change{Combining our interview findings and a systematic literature review of prior implications on inclusive avatar design (e.g., \cite{zhang2022, zhang2023, kelly2023, assets_24})}, we derived 20 \change{initial} 
guidelines that cover five 
aspects of avatar design, including avatar body appearance (G1), body dynamics (G2), assistive technology design (G3), peripherals around avatars (G4), and customization controls in the avatar interface (G5). 
We then validated the guidelines by conducting a heuristic evaluation with 10 VR developers and designers. 
Our evaluation results confirmed that the guidelines comprehensively cover a broad range of disability expression methods through diverse avatar design aspects, are applicable to various avatar platforms, and are actionable to implement with concrete avatar examples. \change{Based on developers and designers' feedback, we further refined the guidelines to make them more accessible and actionable to industry practitioners (e.g., adding applicable use cases, defined the customization scope). As a result, we ended up with a set of 17 finalized guidelines upon practitioners' feedback.} 

Our paper presents two major contributions. First, \change{we contribute a centralized, comprehensive set of guidelines with concrete evidence and implementation examples to support VR practitioners in creating appropriate avatars for PWD.} 
Second, we validated \change{and refined} our guidelines with VR practitioners and confirmed the coverage, applicability, and actionability of the guidelines. 
To ensure practitioners can easily adopt, share, and update the guidelines, we have open-sourced the full version of the guidelines on GitHub\footnote{Inclusive Avatar Guidelines and Library. \url{https://github.com/MadisonAbilityLab/Inclusive-Avatar-Guidelines-and-Library}.}. 
\section{Related Work}

In recent years, social VR has emerged as a promising digital social space, where users could meet and interact via embodied avatars \cite{Tham_2018_understanding, maloney_2020_sleep}. Unlike 2D virtual worlds where user can only control avatars by mouse and keyboard, 
social VR offers more immersive experiences through full-body tracking avatars \cite{Freeman2020, Freeman2021, Zamanifard_2019_togetherness, Caserman_2019_real_time}. Prior work has summarized the unique characteristics of social VR avatars, including rendering full-body movements and gestures in real time \cite{Zamanifard_2019_togetherness, Sra_2018_your_place}, affording vivid spatial and temporal experiences in high-fidelity 3D virtual spaces \cite{McVeigh_2019_shaping, McVeigh_2018_what}, and mediating both verbal and non-verbal communications through facial expressions and body language \cite{maloney_nonverbal, Moustafa_2018_longitudinal}.
\change{These characteristics create a strong sense of embodiment. As a result, many users see avatars as their proxy in the virtual social space and thus curate the avatar design for self-representation \cite{Freeman2021, Morris_2023_women, freeman_2022_noncis}. In this section, we summarize prior literature on avatar-based self-representation, misrepresentation of disability in social media, and  guideline generation and evaluation methods to contextualize our research.} 

\subsection{Avatars and Self-representation in Social VR}
A growing body of HCI work has explored the avatar-based self-representation for underrepresented user groups, such as racial minorities ~\cite{VRharass2019, freeman2022disturbing, Freeman2021}, women~\cite{Morris_2023_women, Sadeh-Sharvit_2021_sexual}, LGBTQ ~\cite{freeman2022disturbing, queeractingout2022, freeman_2022_noncis}, and older adults ~\cite{Baker2021elder}. For example, Freeman et al. ~\cite{freeman_2022_noncis} interviewed 59 non-cisgender users to explore their avatar choices in social VR, revealing that they used avatars with different genders to signify t heir flexible and fluid gender identities. Morris et al. ~\cite{Morris_2023_women} studied avatar representations at the intersection of age and gender. By interviewing ten women in midlife, they found participants desired to use nuanced avatar options with a gradation colors, textures, and body types for self-expression, avoiding hyper-sexualized or stereotypical portray of midlife women. 

\change{Recently, more research attention has been drawn to PWD in social VR, investigating their avatar representation preferences ~\cite{zhang2022, zhang2023, isit_assets24, kelly2023, assets_24, chronic_pain_gualano_2024, ribeiro2024towards}, with four papers ~\cite{zhang2022, kelly2023, assets_24, zhang2023} deriving design implications for inclusive avatar design.} For example, Zhang et al. ~\cite{zhang2022} interviewed 19 \change{people with sensory disabilities, including people who are blind or low vision people and people who are d/Deaf or hard of hearing, to understand their avatar preferences} and revealed a spectrum of disability expression strategies, such as accurately reflecting all disabilities, or disclosing only selective disabilities.
\change{Three design implications were derived, including offering diverse assistive technology options, supporting invisible disability representation through awareness-building accessories, and avoiding misuse of disability features.} 

Moreover, Gualano at al. \cite{assets_24} interviewed 15 people with invisible disabilities (e.g., ADHD, dyslexia), finding that they wished to leverage the embodied characteristics of avatars (e.g., facial expressions and body language) to dynamically present their disability conditions. \change{The research therefore discussed four design implications, such as creating more options for multi-modal communication and enabling dynamic presentation of fluctuating disability conditions.} 
Beyond focusing merely on disability identity, Mack et al. ~\cite{kelly2023} considered the intersectional identities, such as disability intersecting with race and ethnicity. They revealed that participants with intersecting identities may face conflict in presenting these identities and had to make trade-offs when deciding which identities to present. \change{They derived seven design implications, such as allowing customization of body sizes and enabling users to save and switch between multiple versions of avatars.} 

While prior research has explored the self-representation preferences of PWD and derived corresponding implications, \change{these work focused on different aspects and the design implications were dispersed across multiple publications and venues \cite{zhang2022, assets_24, kelly2023}, being difficult for practitioners to collect and apply comprehensively}. \change{There is also a lack of systematic evaluation on their usability and actionability.}  
\change{Centralized and systematically-validated design guidelines are needed to enable easy adoption and dissemination.} 


\subsection{Stigma and Misrepresentation of PWD in the Media}
\change{Improper disability representation can lead to misconception and perpetuate stigma} 
\cite{kelly_AI24, ippolito2020misrepresentation, hall1997representation, young2014imnot}. PWD are seldom represented in the media \cite{darke2004changing}, and when they do appear, they are often misportrayed as objects of pity, superheroes who have accomplished great feats, or people in need of charity and help \cite{garlandthomson2002politics, quayson2007aesthetic, young2014imnot}. Prior work has studied the misrepresentation of disability in various forms of digital media \cite{garlandthomson2002politics, elliott1982media, kelly_AI24}. For example, Garland-Thomson ~\cite{garlandthomson2002politics} analyzed the visual rhetoric of disability in popular photography and found many of them propagated PWD as being pitiful. Leonard \cite{elliott1982media} revealed that television stigmatized PWD by portraying them as being predominately from lower social classes and unemployed. With the advent of AI technology, Mack et al. \cite{kelly_AI24} explored the disability representation in text-to-image generative AI systems, highlighting that the AI-generated images perpetuated the misconception of PWD as primarily using wheelchairs, being sad and lonely, incapable, and inactive.

Compared to conventional 2D media like images or videos, misrepresentation of disability via avatars in the 3D social VR space can be more harmful to PWD due to its highly embodied and interactive nature \cite{blackwell2019harassment, VRharass2019, freeman2022disturbing}. \change{Researchers started to investigate the experiences of using avatars with disability representation features in social VR \cite{zhang2023, isit_assets24}.} For example, Zhang et al. \cite{zhang2023} conducted a two-week diary study with 10 PWD to compare their experiences when using avatars with and without 
disability signifiers (e.g., an avatar on a virtual wheelchair), \change{revealing different types of harassment targeting PWD (e.g., physical harassment, mimicking disability). They further highlighted the misunderstanding between PWD and other social VR users---while PWD used avatars to present their disability identity, others misinterpreted these avatars as trolling or meme avatars, leading to offensive social behaviors.}
To avoid misinterpretation and harassment, they suggested improving the quality and aesthetics of disability features on avatars. 
\change{Similarly, Angerbauer et al. \cite{isit_assets24} conducted a one-week diary study to understand the experiences of 26 PWD when using avatars with disability signifiers, including both visible and invisible ones (e.g., avatar with a sunflower patch). They found that participants experienced both positive and negative social interactions regardless the visibility of their disability signifiers.}
These works indicated that, even with the goal of designing signifiers for positive disability representation, many design choices can still result in negative user experiences. \change{This emphasized the needs for inclusive avatar design guidelines generated with the PWD community to effectively assist VR practitioners in creating avatars that represent disabilities properly.} 

\subsection{Guideline Generation and Evaluation} 
Design guidelines are important resources that support designers and developers to create positive user experiences. 
\change{Mature fields usually have well-established guidelines for practitioners to follow, such as 2D interface design guidelines~\cite{benyon2014designing}, haptic interaction guidelines \cite{sjostrom2002non}, human-AI interaction guidelines \cite{human_ai_guidelines}, and accessibility guidelines for different platforms \cite{w3c_wcag21_2018, w3c_mobile_accessibility_2015}. These guidelines are often derived from systematic literature review \cite{human_ai_guidelines, wei_2018, Mueller_2014, webdyslexia2012}.} 
For example, Amershi et al. \cite{human_ai_guidelines} consolidated 18 human-AI interaction guidelines from 
both academia publications and industry sources. Muller et al. \cite{Mueller_2014} developed a set of movement-based game design guidelines by reviewing existing literature and incorporating insights from 20 years of working experiences in the gaming industry. Santana et al. \cite{webdyslexia2012} surveyed the state of the art on dyslexia and web accessibility and compiled 41 guidelines that support people with dyslexia in using website. \change{Following a similar method, we seek to derive inclusive avatar design guidelines by reviewing prior literature~\cite{zhang2022, zhang2023, isit_assets24, kelly2023, assets_24, chronic_pain_gualano_2024, ribeiro2024towards}. However, as research on avatar representation for PWD is still nascent, with publications emerging only in recent years (mainly between 2022 and 2024), we further interviewed 60 people with diverse disabilities to supplement prior literature, ensuring comprehensive coverage of our guidelines.}

\change{Guideline evaluation is critical to ensure practitioners to properly act upon and apply the guidelines in development~\cite{lotfi2022taxonomy}.} A commonly used guideline evaluation method is \textit{heuristic evaluation}~\cite{nielsen1990heuristic}, where evaluators examine an interface for the application and violations of a given set of usability heuristics \cite{human_ai_guidelines, wei_2018}. For example, to improve and validate their speech interface guidelines, Wei et al. \cite{wei_2018} asked eight experts to follow heuristic evaluation and examine three speech-based smart devices with the guidelines. Amershi et al. \cite{human_ai_guidelines} adopted similar method to evaluate their human-AI interaction guidelines. \change{They modified the heuristic evaluation method to focus on evaluating the guidelines rather than an interface.} 
Specifically, participants first evaluated an AI-based interface with the proposed guidelines and then reflected on the guideline usability. 
We adopted the similar modified heuristic evaluation method to evaluate and refine our design guidelines for inclusive avatars.

\begin{table*}
    \centering
    \begin{tabular}{
    >{\raggedright\arraybackslash}p{1.55cm} c|
    >{\raggedright\arraybackslash}p{0.7cm} c|
    >{\raggedright\arraybackslash}p{3.4cm} c|
    >{\raggedright\arraybackslash}p{3.3cm} c|
    >{\raggedright\arraybackslash}p{1.56cm} c}
        \toprule
        \multicolumn{2}{c}{\textbf{Gender}} & \multicolumn{2}{c}{\textbf{Age}} & \multicolumn{2}{c}{\textbf{Race}} & \multicolumn{2}{c}{\textbf{Disability}} & \multicolumn{2}{c}{\textbf{Avatar Experience}} \\ \toprule
        Female & 53.3\% & 18-24 & 28.3\% & White & 53.3\% & Mobility disability & 28.3\% & Meta Avatar & 53.3\% \\ 
        Male & 38.3\% & 25-34 & 33.3\% & Black or African American & 20.0\% & Mental health conditions & 20.0\% & VRChat & 31.7\% \\ 
        Non-binary & 8.30\% & 35-44 & 16.7\% & Asian & 11.7\% & Sensory disability & 11.7\% & Minecraft & 15.0\% \\ 
        & & 45-54 & 8.30\% & Hispanic or Latino & 6.70\% & Neurodivergence & 11.6\%  & Snapchat & 15.0\% \\ 
        & & 55+ & 13.3\% & Mixed race & 6.70\% & Chronic illness & 1.70 \% & AltspaceVR & 10.0\% \\ 
        & & & & American Indian or Alaska Native & 1.70\% & Multiple disabilities & 26.7\% & Rec Room & 6.70\% \\
        \bottomrule
    \end{tabular}
    \Description{The table is titled "Table 1. Participants’ demographics distribution (in percentage)." It presents the demographic breakdown of study participants, including categories for gender, age, race, disabilities, and avatar experiences. The percentages for each category are provided.}
    \caption{Participants' demographics distribution (in percentage)
    , including gender, age, race and ethnicity, self-reported disabilities, and avatar experiences.} 
\label{tab:pwd_demographics}
\end{table*}

\section{Guideline Generation}

\change{To derive a comprehensive set of design guidelines, we integrated prior design implications for inclusive avatars \cite{zhang2022, zhang2023, assets_24, chronic_pain_gualano_2024} 
with findings from a supplementary interview study with 60 PWD.} 
The interview study was approved by the Institutional Review Board (IRB) at our university. 


\subsection{\cameraready{Systematic Literature Review} of Inclusive Avatar Designs} \label{literature_review}

\change{We first conducted a \cameraready{systematic literature review} on avatar design for disability representation to (1) summarize prior design implications to integrate into our guidelines and (2) inform our interview study, thus covering various perspectives of avatar design to derive comprehensive guidelines. 
}


To collect prior work thoroughly, we searched on Google Scholar and representative HCI and AR/VR venues (i.e., ACM CHI, CSCW, IEEE VR, ISMAR), using keywords: ``inclusive avatars,'' ``avatar design,'' ``avatar representation,'' combined with ``people with disabilities,'' ``disability representation,'' and ``social virtual realities (or VR).'' We filtered the searched results by manually examining the full text of each paper and removing irrelevant and repetitive papers. \change{In the filtering process, we also identified papers discussing avatar designs for other minoritized identities, such as race \cite{Lee2014} and gender \cite{Morris_2023_women}. We included these papers in our review to ensure our interview protocol covers the intersectionality of different identities.}

As a result, we collected a list of \change{19} papers that focused on inclusive avatar designs and identified five key design aspects, including: (1) visual representation of avatars' physical appearance ~\cite{zhang2022, kelly2023, Freeman2020, DavisStanovsek2021, Lee2014, lee_2021, McArthur_2015, McArthur2014, do2023valid, Weidner_2023, Morris_2023_women};
(2) facial expressions and body movements \change{~\cite{assets_24, Gualano_2023, HerreraBailenson2021, Weidner_2023, kopf_2023, chronic_pain_gualano_2024}};
(3) design of assistive technology and their interactions with avatars ~\cite{zhang2022, ParkKim2022, zhang2023, isit_assets24};
(4) avatar's outfits, accessories, and peripheral design elements ~\cite{zhang2022, zhang2023, DavisStanovsek2021};
and (5) design of avatars representing multiple and intersecting identities ~\cite{kelly2023}.
\change{We thus structured our interview protocol based on these five design aspects but encouraged participants to expand to other aspects if they can think of any.}

\change{Among the 19 papers, four papers \cite{zhang2022, kelly2023, assets_24, zhang2023} derived design implications on inclusive avatars for PWD, including seven implications from Mack et al. \cite{kelly2023}, four from Gualano et al. \cite{assets_24}, three from Zhang et al. \cite{zhang2022}, and one from Zhang et al. \cite{zhang2023}. We thoroughly collected these design implications and cross-referenced them with our interview findings, ensuring the comprehensive coverage of derived guidelines.} 

\subsection{Participants}
\change{We further conducted a semi-structured interview with 60 participants with diverse disabilities to supplement prior insights.} To ensure a wide coverage, 
we broadly recruited people who identified as PWD without restricting the disability types. We recruited via multiple channels, including the mailing lists of non-profit disability organizations (e.g., the United Spinal Association), disability and VR communities on mainstream social media platforms (e.g., DisabiliTEA on Discord, r/amputee on Reddit), referrals from recruited participants, and our university’s research forum. Interested participants completed a screening survey with their age, disability conditions, and general social VR and avatar experiences. Eligible participants must (1) be over 18 years old, (2) identify as having disabilities, and (3) have experience using avatars in social virtual worlds and VR platforms. We limited our recruitment to individuals who spoke English. If selected, participants were asked to complete an oral consent at the beginning of the study.

We recruited 60 participants, with ages ranging from 18 to 71 ($mean$ = 34, $SD$ = 13). Our participants covered a comprehensive set of disabilities, with a total of 26 unique disabilities in six categories, including mobility disabilities (e.g., amputee, cerebral palsy), mental health conditions (e.g., anxiety, depression), sensory disabilities (e.g., blind or low vision), neurodivergence (e.g., autism, ADHD), chronic illness conditions (e.g., Epidermal Nevus Syndromes), and multiple disabilities. 
These categories encompassed both visible and invisible disabilities, \change{presenting a wide coverage on disability characteristics and social implications}. \change{Notably, sixteen participants experienced more than one type of disability, offering insights into managing multiple disabilities in avatar design.} 

Moreover, \change{as disability may intersect with other identities} (e.g., gender, race, and ethnicity) and \change{result in unique identity representation strategies} \cite{Crenshaw1990, kelly2023}, we recruited people with diverse gender and racial identities \change{to discuss potential conflicts and trade-off in presenting multiple identities via avatar design}. As a result, our participants included 32 female, 23 male, and five non-binary. Regarding race, 32 participants identified as White, 12 as Black or African American, seven as Asian, four as Hispanic or Latino, four as mixed race, and one as American Indian or Alaska Native. 

All participants had created and used avatars in social virtual environments. Their commonly used platforms included \textit{Meta Avatars} \cite{metaavatars}, \textit{VRChat} \cite{vrchat}, \textit{Minecraft} \cite{minecraft}, \textit{Snapchat} \cite{snapchat}, \textit{AltspaceVR} \cite{altspacevr}, and \textit{Rec Room} \cite{recroom}.
Table \ref{tab:pwd_demographics} presents participants' demographic breakdowns from different aspects.

\subsection{Procedure}

The interview was a single-session study conducted in English, lasting approximately 60 minutes. We conducted interviews via Zoom, and participants received \$25/hour compensation upon completion. \change{Our interview included three phases: an introduction phase, an avatar design phase, and a reflection phase.} 

\change{In the introduction phase, we first went through the consent form with participants and obtained their oral consent. We then asked about participants' background information, including their}
demographics (i.e., age, gender, race, and ethnicity), self-reported disability, and experiences with avatars and social VR platforms. 

In \change{the avatar design phase}, we focused on understanding participants' design decisions and disability expression preferences. \change{To better contextualize participants and encourage them to consider avatar design details}, we employed an example-oriented approach by first asking participants to imagine and describe their ideal avatar designs. \change{We then dived into the five design aspects informed by our \cameraready{systematic literature review} (Section \ref{literature_review})---physical appearance, facial expressions and body movements, assistive technology and interactions with it, outfits and accessories, and multiple and intersecting identities---and asked about participants' design choices for that ideal avatar along each aspect, respectively.} 

\change{Specifically for each design aspect,}  
we broke it down into the smallest possible design elements to fully explore the avatar's potential for disability expression. For example, when asking about the physical appearance of avatar, we began from a higher level of general avatar types to the visual details of each component of avatar body, including physical composition, \change{skin color,} body size and shape, the proportion of different body parts, facial features (\change{e.g., appearances of eyes, cheekbones, nose, mouth, forehead, and jawline}), and hairstyles. 
For each design element, we asked participants about how they wanted to design it, followed up with if and how their choices may reflect their disabilities and the rationales. 
\change{For the multiple and intersecting disability aspect, we included a few dedicated questions to ask about how participants would like to represent fluctuating disabilities and multiple identities if they have any, whether they have encountered any conflicts in representing multiple identities, and their coping strategies.}

\change{Lastly, in the reflection phase, we asked participants to discuss other design aspects if they could think of any and whether and how they may express their disability along that aspect. Moreover, we asked about their concerns of expressing disabilities on avatars and preferences for avatar interface design. 
We concluded the study by discussing the general insights on designing avatars for PWD, including what to do and to avoid.}

\subsection{Analysis \& Guideline Generation} \label{analysis_phase1}
Upon participants' consent, we audio-recorded and transcribed all interviews using an automatic transcription service. Two researchers reviewed the transcripts and manually corrected all transcription errors. We conducted thematic analysis \cite{Braun2006Thematic, Clarke2015Thematic} to identify repetitive patterns and themes in the interview data. First, two researchers open-coded ten identical sample transcripts (more than 15\% of the data) independently at the sentence level. We created an initial codebook by discussing and reconciling codes to resolve any differences. Next, the two researchers split up the remaining transcripts and coded them independently based on the initial codebook. Throughout this process, they regularly checked each other's codes to iterate upon and refine the codebook until a complete agreement. Meanwhile, a third researcher oversaw all coding activities to ensure a high-level agreement. As we reached complete agreement for our coding, inter-coder agreement was not necessary~\cite{mcdonald2019reliability}. \change{The final codebook contains 239 codes.}

\change{Our theme generation adopted a mixed analysis method, where we used design recommendations from prior literature ~\cite{zhang2022, zhang2023, kelly2023, assets_24} to inform deductive themes, along with an inductive method to generate new themes and guidelines from the interview data via an affinity diagram \cite{clarke2017thematic}.} 
After the initial themes were identified, researchers cross-referenced the original data, the codebook, and the themes, to make final adjustments, ensuring that all codes fell in the correct themes. 
\section{Description of Initial Guidelines}

Building upon the knowledge from our literature review and interview study, 
we derive 20 design guidelines for inclusive avatar design for PWD. Our guidelines cover a broad range of disability expression methods across five aspects, including avatar appearance (G1), body dynamics (G2), assistive technology design (G3), peripherals around avatars (G4), and customization controls in the avatar interface (G5). 

To ensure actionability, each guideline has three components: a detailed description, quote examples from PWD, and concrete avatar feature examples to demonstrate the guideline implementation. 
Appendix Table \ref{tab:overview_original} 
presents a summarized version of our initial guidelines. 
In this section, we elaborate on each guideline and the rationales, \change{grounded in both prior literature and findings from our interview study}. 


\subsection{Avatar Body Appearance (G1)}
\change{Customizing avatar body appearance is the most common way to express disabilities. But deciding how to represent disabilities via avatar body is challenging, as it involves multiple design considerations (e.g., body compositions, customization of each body part). We derive guidelines to inform suitable avatar body design for PWD.
}

\textbf{\textit{G1.1. Support disability representation in social VR avatars.}} 
All participants desired the options to represent their disabilities via avatars, \change{echoing the findings from prior works \cite{kelly2023, zhang2022, zhang2023}}. As P52 described: \textit{``I think the biggest thing for me is the flexibility and the freedom to choose [how I can represent myself.]}


\change{\textbf{\textit{Guideline}}: Avatar interfaces should allow all users to flexibly express their identities and present their disabilities ~\cite{zhang2022, kelly2023, assets_24}. The disability features should not be blocked behind the paywall ~\cite{kelly2023}.} 


\textbf{\textit{G1.2 Default to full-body avatars to enable diverse disability representation across different body parts.}}
Almost all participants preferred a full-body avatar, \change{echoing prior insights from Mack et al. \cite{kelly2023}}. In our study, more than half of the participants had a disability that affected the below-waist body area, which could only be reflected via full-body avatars. P9 indicated: \textit{``people can only get my whole disability identity with the full body [avatar].''}

In addition, multiple participants (e.g., P2, P34, P52) wanted to express their lived experiences as PWD, which could only be achieved \change{through the behaviors of full-body avatars}. As P34 described: \textit{``I prefer a full-body avatar, because [it shows] how [people with visual impairments] navigate the pathway, how they move the hands and the fingers, and how they read braille.''} 

\change{\textbf{\textit{Guideline}}:} Avatar interfaces should offer full-body avatar options \cite{kelly2023}. Given the dominant preferences for full-body avatars over others (e.g., upper-body only, or head and hands only), we recommend making it the default or the starting avatar template, giving users the maximum flexibility to further customize their avatars as they prefer. 

\textbf{\textit{G1.3 Enable flexible customization of body parts as opposed to using non-adjustable avatar templates.}}
\change{Similar to prior work ~\cite{zhang2022, kelly2023}}, a third of participants (e.g., P3, P37, P52) preferred matching the avatar body with their authentic self, requesting full customization of the avatar body. \change{Moreover, our participants highlighted the need for asymmetric designs of avatar body parts (P4, P9). For example, P9 wanted avatars with different eye appearance to reflect her amblyopia (i.e., lazy eyes): \textit{``[I want] the opportunity to move or articulate the eyeballs to represent my disability, because my right eye is litter lazier than the other one.''}}

\change{\textbf{\textit{Guideline}}: Avatar interfaces should provide PWD sufficient flexibility to customize each avatar body part \cite{kelly2023}. While the customization spans a wide range, the most commonly mentioned body parts to customize include (1) avatar height, (2) body shape, (3) limbs (i.e., number of limbs, length and strength of each limb), and (4) facial features (e.g., mouth shape, eye size). Asymmetrical design options of body parts (e.g., eyes, ears) should also be available, 
such as changing size and direction of each eyeball to reflect disabilities like strabismus.}



\textbf{\textit{G1.4 Prioritize human avatars to emphasize the ``humanity'' rather than the ``disability'' aspect of identity.}}
Approximately half of participants chose \change{\textit{human}} avatars to stress disability as an inherent part of personal identity. Multiple participants (e.g., P9, P31, P46) reported real-life experiences of being degraded and wanted to use human avatars to express that they should be seen as ``a whole person, not the disability'' (P46). Additionally, the human avatars also let users to represent multiple and intersectional identities (e.g., age, gender, and race) all together as an integral human being. As P9 emphasized: \textit{``Humanoid avatars show me as a whole person. I identify as a black woman with a disability, and that's really important when discussing a personal identity, because when describing somebody, you wouldn't just say, ‘Oh, they have a disability,’ instead you would say, ‘Oh, they're non-binary or female, and they're African American or Caucasian, and they have a disability.’ They all go together.''} 

\change{\textbf{\textit{Guideline:}} Social VR applications should offer human avatar options whenever the application theme allows. 
} 

\textbf{\textit{G1.5 Provide non-human avatar options to free users from social stigma in real life.}} 
\remove{Unlike those who authentically indicated disabilities with accurate details,} Eleven participants (e.g., P17, P43, P49) preferred non-human avatars, such as robots or animal characters, to avoid disclosing disabilities and protect themselves from the judgment they often faced in real life. \remove{They perceived social VR as a utopia where they can escape from real life, and avatars in non-humanoid forms, such as robotics or animal avatars, allowed them to avoid the social judgments and norms commonly tied to disabilities.}  
For example, \remove{P45 said: \textit{``[Non-human avatars] hide more of my real disabilities, being the opposite of showing off my disabilities.''}} P49, who identified as neurodivergent, chose a robotic avatar to lower social expectations: \textit{``It feels like people's expectations of how neurotypical I'm gonna seem are lowered if I'm a robot or just a non-human avatar.''} 

\change{\textbf{\textit{Guideline:}} Besides human avatars, avatar interfaces should also provide diverse forms of non-human avatars, empowering PWD to choose the one they relate with flexibly.}

\subsection{Avatar Dynamics: Facial Expressions, Posture, and Body Motion (G2)}
Unlike 2D interfaces, the embodied and multi-modal nature of avatars in social VR enabled PWD to express themselves through diverse approaches beyond static appearance. \change{We derive guidelines based on how PWD leveraged avatar dynamics, such as facial expressions, posture, and body motions, to represent disabilities.} 
 
\textbf{\textit{G2.1 Allow simulation or tracking of disability-related behaviors but only based on user preference.}}
Nine participants (e.g., P7, P47) wanted their avatars to reflect the realistic behaviors caused by disability for a stronger connection (e.g., limp by P18, stumbling by P4). 
However, \change{similar to Gaulano et al. \cite{chronic_pain_gualano_2024}}, eight participants (e.g., P6, P14, P49) were concerned that showing disability-related behaviors would reinforce stigma (e.g., involuntary behaviors like nervous tics). \remove{Participants preferred avatars to reflect behaviors with their controls. For example, P6 didn't want her avatar to reflect disability-related movements at all to avoid misconception, as she explained: \textit{``The way I move authentically is kind of jaggy, and I swerve. People asked me if I'm drunk all the time. So I'd like to go as quickly as I can in a smooth way [...] even though that's not authentic.''}} \change{Participants highlighted the need for controlling what behaviors to track or simulate (e.g., P14, P16, P49).} As P14 indicated: \textit{``I think it would be cool if you could choose to have [the movement to be reflected]. But I also think there is a fine line between inclusion and offensive imitation.''} 

\change{\textbf{\textit{Guideline:}} Users should be able to control the extent of behavior tracking in social VR. With the advance of motion tracking techniques, avatar platforms may disable subtle behavior tracking by default to avoid disrespectful simulation, but allow users to easily adjust the tracking granularity for potential disability expression.}



\textbf{\textit{G2.2 Enable expressive facial animations to deliver invisible status.}}
A third of participants (e.g., P1, P4, P40) desired to express disabilities through avatar facial expressions. This is particularly important for people with invisible disabilities, whose conditions mostly surface through emotions and subtle non-verbal cues. 
Three participants (P4, P40, P51) noted that their disabilities involved rapid fluctuation or contradicting feelings (e.g., bipolar disorder, ADHD), thus preferring avatars to show a spectrum of facial expressions. For example, \change{P51 experienced multiple invisible disabilities (i.e., depression and ADHD) and used different facial expressions to represent different aspects of their disabilities}: \textit{``When representing depression, the facial expression is more sad or in thought. When having ADHD moments, [the avatar] being more excited or manic.''} 

\change{\textbf{\textit{Guideline:}} Avatar platforms should enable diverse facial expressions, allowing PWD to express emotion, portray mental status, and indicate fluctuation of invisible disabilities \cite{assets_24}.  
}

\textbf{\textit{G2.3 Prioritize equitable capability and performance over authentic simulation.}}
Four participants (P7, P14, P15, P52) highlighted that the avatar performance should demonstrate equitable capabilities to other users, not being limited by their disabilities or direct motion tracking. 
For example, P52 mentioned that the moving speed of her avatar walking with limps should not be slower than other avatars: \textit{``I walk with a slight limp, [but] I don't think I need the actual movement [on avatars] to reflect how [fast] I walk. \remove{[Because] when I use games, I see the movement aspect more of a practicality than part of the game [...]} 
I think having a limp would be cool, but I wouldn't want to be slower than [other avatars]. \remove{I wouldn't want to have a maximum speed, because I chose to have a limp earlier in the avatar making process [...]} 
Being able to just keep up with peers’ [avatars], pace-wise, would be the most important thing.''} 

\change{\textbf{\textit{Guideline:}} PWD value equitable and fair interaction experiences more than the authentic disability expression. Therefore, avatar platforms should ensure the same level of capabilities and performance for all avatars, regardless of whether disability features or behaviors are involved.}

\textbf{\textit{G2.4 Leverage avatar posture to demonstrate PWD's lived experiences.}}
In addition to the facial expressions and body movements, five participants (P4, P9, P31, P33, P34) preferred leveraging avatar postures to demonstrate their lived experiences. For example, as a person with low vision, P4 described his unique posture when interacting with others: \textit{``[My] vision is directed at one angle. So my head is turned lightly, because I'm not looking at people directly all the time.''} Representing postures and mannerisms on avatars also help increase awareness and resolve misunderstandings about disabilities, as P34 shared: \textit{``Instead of looking at the person who is speaking, people with visual impairments take their ears nearby to the place where the sound is coming from. This gives some wrong impressions to the people that the visually impaired people have not given attention to the speakers. That is not the real story.''}

\change{\textbf{\textit{Guideline:}} Disabilities can be expressed via avatar posture. Avatar platforms should enable certain posture tracking or simulation (e.g., unique facing directions of individuals with low vision during conversation) to enable authentic disability representation.}


\subsection{Assistive Technology Design (G3)}
Adding assistive technologies to avatars is a key method adopted by PWD for disability disclosure ~\cite{zhang2022, kelly2023}. \change{Beyond prior literature, we revealed key aspects of assistive technologies, such as types, appearances, and relationship to avatars, to guide proper designs.} 

\textbf{\textit{G3.1 Offer various types of assistive technology to cover a wide range of disabilities.}} 
 Like prior work has indicated \cite{zhang2022,kelly2023}, we found that multiple participants (e.g., P18, P33, P39) viewed assistive technologies as part of their body. P39 described the meaning of wheelchairs to wheelchair users: \textit{``\remove{For people in wheelchairs, our wheelchair is an extension of our body.} We view it emotionally as an extension of ourselves, and it gives us our independence.''} \remove{P18 and P33 also reflected that being able to have avatar with assistive technologies they used in daily life made them feel empowering and being included in the social VR.}

\change{\textbf{\textit{Guideline:}}} Avatar interfaces should offer assistive technologies that are commonly used by PWD \cite{zhang2022}. \change{The most desired types of assistive technologies include: (1) mobility aids (e.g., wheelchair, cane, and crutches); (2) prosthetic limbs; (3) visual aids (e.g., white cane, glasses, and guide dog); (4) hearing aids and cochlear implants; and (5) health monitoring devices (e.g., insulin pumps, ventilator, smart watches). 
Practitioners should consider including at least these five categories of assistive technologies in avatar interfaces.
In addition, due to PWD's different technology preferences \cite{kelly_AI24}, we encourage practitioners to offer more than one assistive technology option in each category, for example, including guide dog, white cane, and glasses for visual aids.} 

\textbf{\textit{G3.2 Allow detail customization of assistive technology for personalized disability representation.}}
Eleven participants (e.g., P15, P18, P32) desired to better convey their personalities through assistive technology customization. \change{Echoing prior research ~\cite{zhang2022, kelly2023}, changing the colors of assistive technologies and attaching personalized decorations (e.g., add stickers on wheelchair, P18) are two most preferred customization options.}
\remove{With assistive technology being an extension of the user’s body, being able to have diverse customization options of assistive technologies is as important as customizing the avatar’s appearance, as P39 said: \textit{``Making some more customization in the wheelchair [is] in the same way that you make customization for eye color, nose shape, [and] all those things.''}} However, some participants (e.g., P20, P50) 
\change{desired to see more various styles of assistive technologies, such as a futuristic styled hoverchair (P20).}
\remove{as current avatar platforms offer only limited default choices.}  
\remove{As P52 expressed frustration over the lack of w heelchair variations in social VR avatars: \textit{``[Now] you either have a wheelchair or no wheelchair, but you can't customize the type, shape, or any various add-ons. Like is it motorized [wheelchair]? Is it like a manual one? So I think having the ability to choose what additional features you'd like to add would be really nice.''}} 

\change{\textbf{\textit{Guideline:}} Avatar platforms should allow customizations for assistive technology \cite{kelly2023,zhang2022}. Basic customization options should include adjusting the colors of different assistive technology components and adding decorations (e.g., stickers, logos) to them. More customization could be added based on specific use cases.} 


\textbf{\textit{G3.3 Provide high-quality, authentic simulation of assistive technology to present disability respectfully and avoid misuse.}}
Four participants (P4, P6, P34, P35) preferred high-quality assistive technology simulation with authentic details similar to those in real life. They were concerned that inaccurate assistive technology designs in social VR may depict misleading figures of PWD and lead to misuse, echoing Zhang et al. \cite{zhang2023}. As P35 described: 
\textit{``[I’ve seen] really poor representation [of wheelchairs]. They're usually joke avatars or meme avatars that have wheelchairs.''}

\change{\textbf{\textit{Guideline:}} To avoid misunderstandings or misuse, the assistive technolgy simulation should convey standardized, authentic details of the real-world assistive devices \cite{zhang2023}, regardless the overall avatar style. For example, the design of a white cane should show the details of tip and follow its standardized color selection, no matter the design style is photorealistic or cartoon. 
We recommend practitioners to model assistive technologies by following their established design standards, such as design guidelines for white canes \cite{who_white_canes}, wheelchairs \cite{russotti_ansi_wheelchairs}, and hearing devices \cite{ecfr_800_30}.}
\remove{Their designs should be high in quality and contain sufficient details, so that users can tell these avatars were invested with great efforts, aiming for identity representation instead of trolling (P6, P34). For example, the design of a white cane for people with low vision should show the details of tip and follow the standardized color selection for such walking aids (P34).} 

\textbf{\textit{G3.4 Focus on simulating assistive technology that empower PWD rather than highlighting their challenges.}} 
Eight participants (e.g., P18, P39) only wanted to add assistive technology features that can demonstrate their independence instead of challenges, (e.g., hospital wheelchair vs. power wheelchair). 
For example, P18 found media often misrepresented PWD by showing them sitting in a hospital-style wheelchair that requires others' assistance to move: \textit{``Most of the representations we see in fiction, video games and TV, they always use hospital chairs, which are not practical. No actual disabled person uses a hospital chair in real life, which has armrests and big push handles, because it's built for somebody to push you. However, a manual wheelchair is designed for you to push yourself.''}
\remove{Participants wanted to correct the media misrepresentation by showing how they can achieve independence through the use of assistive technology. Taking the wheelchairs as examples, P6 noted the power chair should have a joystick to show the user can move independently; and P39 strongly preferred manual wheelchair without any pushable handles to demonstrate the self-independence: \textit{``It's important to me that it doesn't look like I'm ready to be pushed by someone else. I'm stating that independence [achieved through wheelchair]. I'm solidly myself, and I don't need another person. This is a big deal in our community [...] we’re not going to want push handles.''}}

\change{\textbf{\textit{Guideline:}} When determining what assistive technology features to offer, practitioners should only select assistive or medical devices that can be easily controlled by PWD to demonstrate their capability (e.g., manual wheelchair, cane) and leave out the ones that PWD cannot independently use or the ones that highlight their challenges (e.g., hospital-style wheelchair, bedridden avatars).}

\textbf{\textit{G3.5 Demonstrate the liveliness of PWD through dynamic interactions with assistive technology.}}
In addition to the visual details, five participants (P4, P6, P18, P39, P44) \change{wanted their avatars to actively interact with the assistive technologies, such as rolling their manual wheelchair (P18) or sweeping their cane (P34) when moving, to demonstrate their capability and liveliness}. 
As P34 mentioned: \textit{``While my avatar is walking, the tip of white cane should be moving back and forth like a pendulum motion.''}
\remove{\textit{``I think having the option to roll [wheelchair] would be good. I’ve seen some 3D models of wheelchair users in video games, and their arms don't move while they're rolling, which is really weird to me. Because I push myself with my hands.''}} 

\change{\textbf{\textit{Guideline:}} Beyond providing assistive technology options, social VR platforms should enable suitable interactions between avatars and assistive technology. The interactions should authentically reflect PWD's real-world usage of their assistive technology, such as how a blind user sweeps their cane, or how a wheelchair user moves their arms to control their wheelchair.  
}

\textbf{\textit{G3.6 Avoid overshadowing the avatar body with assistive technology.}}
Seven participants (e.g., P9, P18, P57) demanded to flexibly adjust the size of assistive technologies to fit their avatar body. \change{They emphasized that, while expressing disabilities, the image curation should focus on the whole avatar rather than just the assistive technology. As P39 highlighted: \textit{``The wheelchair is not the focus of the image; [rather,] the focus is on the avatar having a good time.''}} \change{P18 recalled their 
prior experience of being overshadowed by the assistive technology}: \textit{``
\remove{I think that having the option to actually make the chair larger or smaller, depending on how large or small your avatar is, is a good detail. Because sometimes wheelchairs don't fit you.}I have encountered 3d models where the wheelchair is so big and the person sitting in it is so small, and it just doesn't look right.''}

\change{\textbf{\textit{Guideline:}} The size of assistive technology should not dominate the avatar body but rather fit the body size. Avatar platforms should automatically match the assistive technology model to different avatar body sizes, and allow users to adjust the size of assistive technology to achieve the preferred avatar-aid ratio. The combination of avatar and assistive technology should also be seamless without affecting the quality and aesthetics of the original avatar ~\cite{kelly2023}.}

\subsection{Peripherals around Avatars (G4)}
Beyond the design of avatars, the peripheral space around them can also be leveraged for disability expression. We explored this new design space and identified design guidelines.

\textbf{\textit{G4.1 Provide suitable icons, logos, and slogans that represent disability communities.}}
Sixteen participants (e.g., P5, P37, P58) desired 
representative icons, logos, or slogans of disability communities for identity expressions, and they wanted to creatively attach these symbols to a variety of places, such as on avatar's clothing (P53, P56), accessories (P13, P54), or even the space surrounding the avatars (P14, P47). \change{This confirm previous insights that disability-related symbols can help PWD educate other users and raise awareness in the social VR space \cite{zhang2022, assets_24}.} 
\remove{For example, multiple participants mentioned using the rainbow infinity icon to represent the autism community (e.g., P1, P46, P47), zebra printing for rare disease (P56), and sunflowers that symbolize interac tion invisible disabilities community (P5, P14). For example, P14 planned to attach a sunflower yard in the background of her avatar to symbolize her invisible disabilities. 
Another participant, P3, would like to have disability community icons on the avatar's T-shirt to show community pride.} 

\change{\textbf{\textit{Guideline:}} Awareness-building items (e.g., logos, slogans) should be provided, allowing users to attach them to various areas on or around the avatars \cite{assets_24, zhang2022}. 
Some widely recognized and preferred symbols that represent different disabilities for practitioners to refer to include (1) the rainbow infinity symbol that represents the autism community \cite{assets_24, rainbow_infinity_symbol}, (2) the sunflower that represents hidden disabilities \cite{isit_assets24, hidden_disability_sunflower}, (3) the disability pride flag \cite{disability_pride_flag}, (4) the spoons, symbolizing spoon theory for people with chronic illness \cite{kelly2023, assets_24}, and (5) the zebra symbols for rare diseases \cite{assets_24, Gualano_2023}. 
}

\textbf{\textit{G4.2 Leverage spaces beyond the avatar body to present disabilities.}}
Eight participants (e.g., P37, P43, P46) wanted to express disabilities more creatively and flexibly through the space behind the avatar body. \change{This is especially favored by people with invisible disabilities, as it helps visualize PWD's mental conditions. For example, P43 wanted a visual indicator of a cloudy and rainy background to symbolize her depression and anti-social mode at the moment. This echoes prior implications that an avatar's background can provide contexts into PWD's experiences \cite{kelly2023, assets_24}.}
\remove{P47 would like to add a variation of battery symbols over the avatar's head, which would change levels based on her energy: \textit{``My energy levels can fluctuate just a lot. Someday, I may have a little bit of energy, and the next day I may have a lot of energy, and that could actually change within a matter of hours. So the idea that I have is a battery symbol that I could adjust the battery level shown on that to show you how much energy that I have to spend. It's a signal to my friends that ‘hey, my battery's low, I may sound really tired right? I'm okay, I just have low energy.’ [Other times] I could turn my battery all the way up and be like, ‘Hey, let's see, we can do something a little bit more active.’''}}

\change{\textbf{\textit{Guideline:}} When designing avatars, practitioners should consider leveraging avatar's peripheral space to enable users to better express their status, especially for individuals with invisible disabilities. Some design examples include a weather background to indicate mood and a battery sign to indicate energy level \cite{assets_24}. 
}

\subsection{Design of Avatar Customization and Control Interface (G5)}
\change{The usability and accessibility of avatar interfaces can significantly impact PWD's avatar customization experiences. We thus identified guidelines for avatar customization and control interfaces to enable smooth avatar curation for PWD.}


\textbf{\textit{G5.1 Distribute disability features across the entire avatar interface rather than gathering them in a specialized category.}}
Five participants (P18, P32, P44, P49, P57) strongly preferred embedding the disability-related features naturally into different categories of the avatar interfaces \change{(e.g., asymmetrical eyes under the eye category, amputation under the body category), as opposed to collecting them in a specialized category for PWD, which marginalized them by ``setting PWD apart from other users'' (P32)}. 
\remove{P49 suggested developers and designers to treat the disability-related features in the same way as any other avatar features in the interface: \textit{``Just treating them as neutral instead of either a burden to have to design or something you get to feel really special for designing''}.}

\change{\textbf{\textit{Guideline:}} Avatar features for disability expression should be treated in the same way as other avatar features. In avatar interfaces, disability-related features should be properly distributed in their corresponding categories. There should not be a specialized category for PWD. 
 For example, assistive technologies should be included in the accessory category rather than a separate assistive technology category. }

\textbf{\textit{G5.2 Use continuous controls for high-granularity customization.}}
Eight participants (e.g., P16, P37, P49) believed that the control components in avatar interfaces can largely affect their customization flexibility. To accurately represent their disabilities, participants preferred continuous control methods (e.g., a slider) over discrete options (e.g., binary switches, drop-down menu with limited options). As P47 said: \textit{``[I prefer] the sliding scale. You can really change [the length of the limb] to a very particular level.''}
\remove{\textit{``It's better to have a spectrum of choices, or even a slider-like for people to change your nuanced level.''} Since disability representation was a spectrum (P16, P48), input controls that offered a continuous range of options, such as sliders and knobs, were preferred in the avatar customization interface. As P47 said: \textit{``[I prefer] the sliding scale. You can really change it on a very particular level.''}; P48 also agreed that \textit{``slider is better than binary options.''}}

\change{\textbf{\textit{Guideline:}} Avatar interfaces should adopt input controls that offer a continuous range of options to enable flexible customization. This could be widely applied to a variety of design attributes, such as the size and shape of multiple avatar body parts.}

\textbf{\textit{G5.3 Offer an easy control to turn on/off or switch between disability features.}} \label{g5.3}
Twelve participants (e.g., P11, P32, P57) noted that they didn't want to always disclose their disability identities in social VR. Instead, disability representations were often context-dependent \change{\cite{zhang2022, kelly2023, assets_24}}. 
For example, P39 didn't want to use avatars on wheelchair when surrounded by strangers or in unfamiliar VR worlds. \change{Moreover, people with multiple disabilities or fluctuated status  also need a fast and easy control to switch between different avatars (e.g., avatars with different facial expression in G2.2) or adjust status indicators (e.g., the weather background in G4.2) to flexibly update their disability expression based on contexts.}  
\remove{\textit{``Although I have a disability and I'm comfortable with it, it is not the most important thing to me. Sometimes I might not want to lead with [my disability], especially when you have physical disabilities that people can see [but] you have no control over how people see you right away.''}}


\change{\textbf{\textit{Guideline:}} Interface should provide easy-to-access shortcut control that enable users to conduct \textit{ad-hoc} avatar updates. Important control functions include: (1) toggling on and off the disability-related features \cite{assets_24}; (2) switching between different saved avatars \cite{kelly2023}; and (3) updating status for fluctuating conditions ~\cite{assets_24}. 
}

\section{Guideline Evaluation}

To evaluate the coverage, applicability, and actionability and further refine the guidelines, we conducted a heuristic evaluation study with 10 VR experts to solicit professional feedback.

\subsection{Participants}
We recruited 10 participants with professional experiences in AR/VR development with ages ranging from 24 to 52 years old ($mean$ = 33.5, $SD$ = 10.3). To ensure participants have sufficient avatar development experiences and valuable insights, we focused on recruiting AR/VR developers and designers with at least six months of full-time work experience. We used multiple channels for recruitment, including emailing AR/VR technology companies, posting recruitment messages on online developer forums (e.g., Meta Developer Forum) and business-focused social media platforms (e.g., LinkedIn), and referrals from participants. Interested participants would fill out a survey, asking about their age, current position, years of experiences as AR/VR practitioners, companies they have worked at, and experiences with avatar development. 

Our participants showed a variety of professional experiences with VR development. Five of them had over five years of experience, and six had held leadership roles within their teams. Their senior experiences offered strategic insights about guideline implementations. In addition, all participants had experiences with avatar development in 3D environments, including avatar modeling (e.g., EP1, EP3), 3D texturing (EP8, EP9), rigging (e.g., EP5, EP7), and designing avatar interactions in the social settings (EP1, EP2). EP1 specifically designed and developed avatars for people with depression. Moreover, six participants were employed at medium to large technology companies, while four participants worked at startups or local VR studios. Such diverse backgrounds of participants enabled us to gain feedback from different perspectives and improve our guidelines comprehensively. We detailed participants' demographics in Table \ref{tab:experts_demographics}.

\begin{table*}[ht]
\centering
\renewcommand{\arraystretch}{1} 
\begin{tabular}{
>{\raggedright\arraybackslash}p{1cm} 
>{\raggedright\arraybackslash}p{1.5cm} 
>{\raggedright\arraybackslash}p{5.2cm} 
>{\raggedright\arraybackslash}p{2.3cm} 
>{\raggedright\arraybackslash}p{3.5cm} 
>{\raggedright\arraybackslash}p{2.2cm}}
\toprule
\textbf{ID} & \textbf{Age/ Gender} & \textbf{Position} & \textbf{Years of Experience} & \textbf{Company size} & \textbf{Evaluated Social VR App} \\
\midrule
EP1  & 27/M & Lead XR design engineer & 3 years & 2-10 employees & Rec Room \\
EP2  & 25/M & AR/VR developer & 6 months & 10K+ employees & VRChat \\
EP3  & 27/M & Unity developer & 5 years & 11-50 employees & Rec Room \\
EP4  & 26/M & AR/VR game developer & 2 years & 501-1K employees & Meta Horizon \\
EP5  & 29/M & Lead VR developer and designer & 1.5 years & 2-10 employees & Roblox \\
EP6  & 40/M & Spatial VR developer and designer & 7 years & 1K-5K employees & Multiverse \\
EP7  & 24/M & VR developer & 1 year & 2-10 employees & Multiverse \\
EP8  & 52/F & Senior XR generalist and artist & 10 years & 11-50 employees & Meta Horizon \\
EP9  & 49/M & Lead XR developer and designer & 10 years & 2-10 employees & VRChat \\
EP10 & 36/M & Lead game designer & 8 years & 1K-5K employees & Roblox \\
\bottomrule
\end{tabular}
\Description{The table is titled "Table 2. Expert participants’ demographics in Phase 2." It includes the demographics of expert participants, listing their ID, age, gender, position, years of working experience, company size, and the social VR app they evaluated.}
\caption{Expert participants' demographics in Phase 2, including ID, age, gender, position, years of working experiences, company size, and the social VR app they evaluated.}
\label{tab:experts_demographics}
\end{table*}

\subsection{Procedure}
To evaluate the guidelines, \change{we employed a modified heuristic evaluation method \cite{human_ai_guidelines}}. Heuristic evaluation is a common usability testing method where evaluators examine a user interface for violations based on a set of usability guidelines \cite{nielsen1990heuristic}. With the primary goal of evaluating the design guidelines rather than the interface, and inspired by prior work \cite{human_ai_guidelines}, we modified the heuristic evaluation by asking participants to identify both applications and violations of the guidelines in off-the-shelf social VR apps and reflect on the guidelines themselves.  

\change{Before the study, we emailed the guideline document to each participant, asking them to read through and get familiar with the guidelines.} 
We selected five widely used off-the-shelf social VR apps for participants to review (app selection is detailed in Section ~\ref{sec:appselect}). We randomly assigned one social VR app to each participant and ensured that each app was evaluated by two participants. 

After familiarizing themselves with the guidelines, participants started evaluating their assigned app in their own time. We emailed the study instruction and asked participants to thoroughly explore the avatar customization process of the app, including all design options and the avatar interface. During their explorations, participants filled a survey to evaluate how our guidelines may apply to the app, following the standard heuristic evaluation \cite{nielsen1990heuristic}. The survey took 1 to 2 hours to finish, and participants were given one week to fill out the survey on their own time, with the compensation of \$50 upon completion. We asked participants to video-record their exploration process to ensure the study rigor. 

When we received the participants' survey response, we invited them to a semi-structured interview to go through and discuss the evaluation results. The interview was conducted via Zoom and lasted about 1 hour, with compensation at a rate \$45/hour. We describe the study details of survey and interview below.

\subsubsection{Heuristic Evaluation Survey}
We instructed participants to evaluate their assigned \change{app} based on the guidelines and asked them to fill out a survey asking a series of questions. We implemented the survey in Qualtrics\footnote{Qualtrics. https://www.qualtrics.com/lp/experience-management/.}. We attached the consent form before the survey questions, and participants gave consent by continuing with the survey after reading the consent form.


\change{We modified the standard heuristic evaluation form by adding more questions asking about the guidelines themselves \cite{human_ai_guidelines}.} 
The survey started with the presentation of a guideline, and we first 
asked participants how they interpret it to check any misinterpretations or confusions. To evaluate guideline's applicability on the app, the survey asked participants to determine if the guideline ``apply'' (i.e., is relevant and should be implemented) to the avatars, and if not apply (i.e., irrelevant or out of scope), to explain why. If a participant found that a guideline should apply to the avatars, we further quantify the level of application by asking participants to rate on a 5-point semantic differential scale from ``clearly violated'' (1) to ``clearly applied'' (5). We also asked participants to explain their ratings with identified examples from the evaluated app, so that we got concrete evidence of applications and violations.

In addition, participants answered a set of three questions to assess each guideline's clarity, actionability, and importance on a 5-point Likert scale, with 1 being the worst and 5 being the best. 
Participants also filled an open-ended question to provide rationales for their ratings and other comments about the guidelines.
We asked the same set of questions to evaluate each guideline consistently. We attached the details of survey questions in Appendix \ref{heuristic_survey}.

\subsubsection{Semi-structured Interview} 
After completing the survey, participants did a semi-structured interview with us, which we asked follow-up questions to the survey responses and gathered cross-guidelines feedback. We began the interview by asking participants' background information to better understand their professional experiences, including demographics (age and gender), current position and the company they worked at, years of working experiences as a full-time position, and previous experiences with developing 3D avatars. We then went through the survey response with participants to resolve any confusions or questions they may have about the guidelines, and asked participants to further elaborate their ratings if needed. 
Lastly, to identify any redundancies or conflicts across different guidelines, we asked participants to reflect on the overall guidelines. We also sought feedback about the language and structure of guidelines so that we could resolve any semantic confusion. A detailed protocol could be found \ref{protocol_experts}. 

\subsection{Apparatus: Selection of Mainstream Social VR Apps for Heuristic Evaluation}\label{sec:appselect}
To determine if and how the guidelines manifest in a variety of avatar platforms, we selected a set of commonly used social VR apps for participants to evaluate in the study. We adopted a maximum-variance sampling strategy \cite{maxwell2012qualitative, human_ai_guidelines} to select social VR apps that covered a wide range of avatar types. 

First, we conducted an exhaustive search on three mainstream VR app stores: \textit{Meta Horizon Store}, \textit{Steam}, and \textit{Viveport}. Our search focused on apps available in the United States in English from March to May 2024. We first searched for the keyword ``social'' in each store and identified a total of 527 apps (86 from \textit{Meta Horizon Store}, 234 from \textit{Steam}, and 163 from \textit{Viveport}). To narrow down the scope and focus on apps with embodied social nature, we filtered the results by checking the app descriptions, tags, and thumbnails to exclude the non-social apps (e.g., intense shooting games, sports), resulting in 25 social VR apps. 

Next, four researchers reviewed all apps independently. We adopted a depth-first traverse strategy, clicking through all available buttons and menu items in the avatar interface. During the review, researchers video-recorded and took notes of all avatar options and the interaction process, including avatar types (e.g., realistic human, cartoonish characters, animals), components (e.g., full body, upper body), customization options, and disability representation features. All researchers then discussed the review results to select apps with diverse avatar styles, enabling us to evaluate the guidelines on a broad range of avatar platforms.  
As a result, we selected five social VR apps: \textit{VRChat}, \textit{Meta Horizon Worlds}, \textit{Rec Room}, \textit{Roblox}, and \textit{Multiverse}. \change{We characterized the avatar interface of each app in Appendix Table \ref{tab:vr_platforms}}. 

\subsection{Analysis}
We used descriptive statistical analysis to analyze the survey response. Specifically, we collected participants' Likert scale ratings on four dimensions for each guideline, including level of application, clarity, actionability, and importance. For each dimension, we calculated the mean and standard deviation of ratings to identify guidelines with low mean score or high deviation (meaning that participants had diverse opinions), which may indicate areas for improvement. We also counted the total number of examples of applications and violations identified by participants, understanding how well our guidelines were applied or implemented in mainstream social VR apps. 

We analyzed participants' open-ended responses via thematic analysis \cite{Braun2006Thematic}, similar to the method in Section \ref{analysis_phase1}. Three representative transcripts were selected as samples and coded independently by three researchers. Researchers discussed code discrepancies and created the initial codebook upon agreement. We then split up the remaining transcripts for individual coding. When new codes emerged, researchers discussed with each other and updated the codebook upon agreement. \change{The final codebook contained 137 codes. We categorized all the codes in to high-level themes using affinity diagram and achieved five themes with 17 sub-themes.}


\subsection{Findings}
Our study resulted in both quantitative and qualitative data that reflected VR experts' feedback. Below, we first report quantitative results from the survey responses, providing an overview of participants' assessments on our guidelines. We then discuss the qualitative data to uncover their considerations and suggestions. 

\subsubsection{\change{Application Level of the Guidelines}.}
Across the five social VR apps, participants identified 109 violations, 47 applications, and 38 neutrals (i.e., with a rating of three in level of application). 
Each participant identified at least one example of application or violation for each guideline, suggesting the relevance of our guidelines \cite{human_ai_guidelines}. By analyzing the mean rating of application level, we found 16 out of 20 guidelines had a mean rating below three. Figure \ref{fig:application} in Appendix showed the mean application level score of each guideline. Participants reported that the evaluated apps either lacked any customization options (e.g., \textit{Multiverse}) or the provided customization did not fully address the needs of disability representation (e.g., \textit{Horizon Worlds}, \textit{Roblox}, \textit{Rec Room}). For example, EP1 noted that \textit{Rec Room}'s avatar customization included a slider for adjusting eye sizes in line with G5.2 (i.e., use continuous controls for high-granularity customization), but it only allowed symmetrical changes, violating G1.3 (i.e., enable flexible customization of body parts as opposed to non-adjustable avatar templates). Another typical violation was lack of access to disability features (e.g., \textit{Rec Room}, \textit{VRChat}, \textit{Roblox}). For instance, EP5 found that \textit{Roblox} had a paywall in accessing the wheelchair options. These violations \change{highlighted the lack of inclusion in mainstream social VR apps.}

\subsubsection{Clarity, Actionability, and Importance of the Guidelines.} 
In addition to evaluating the social VR apps with the guidelines, participants also reflected on the guidelines themselves. In general, participants found our guidelines clear and easy to understand. Most guidelines had mean clarity ratings above four (Figure \ref{fig:clarity} in Appendix), 
except for G1.4 ($mean =$ 3.65; i.e., prioritizing human avatars), where \change{five} participants \change{(e.g., EP4, EP5, EP7)} 
were \change{unsure the meaning} of ``emphasize the `humanity' rather than disability,'' \change{wondering what could be a counterexample.} 

 Participants also found the majority of guidelines actionable \change{as each of them came with concrete implementation examples}. 
 Eighteen guidelines had mean actionability ratings above three (Figure \ref{fig:actionability}), with two exceptions of G2.4 ($mean =$ 2.90; i.e., leverage avatar posture to demonstrate PWD's lived experiences) and G3.5 ($mean =$ 2.80; i.e., demonstrate the liveliness of PWD through dynamic interactions with assistive technology). Although participants found these two guidelines leveraged the unique characteristics of embodied avatars, \change{seven of them (e.g., EP1, EP4, EP7, EP8)} 
were unsure about how to implement them without considering the capabilities and limitations of VR devices (e.g., body tracking). \change{We explain in detail in Section \ref{improvements}.} 

Regarding the importance, all guidelines were rated at or above three on average (Figure \ref{fig:importance}), highlighting the importance of our guidelines. \change{We dive into the rationales in Section \ref{motivations}.}


\subsubsection{Motivations to Implement the Guidelines.} \label{motivations}
Eight participants (e.g., EP1, EP3, EP8, EP10) 
praised the guidelines for providing comprehensive and novel suggestions for developing social VR avatars. They found the guidelines not only covered diverse user groups (EP8, EP9) but also inspired design ideas for under-explored avatar features, such as avatar peripherals and dynamics (EP1). \change{We elaborate on the factors that drive participants' motivations}. 

\change{\textbf{\textit{Increased Trust and Values by Engaging Disability Community.}}} Participants were particularly impressed by our inclusion of 60 PWD in generating the guidelines. As developers and designers, they highly valued the direct input from representative users and wanted to incorporate the community's voice in application development. However, this was often hard to achieve due to the limited resources (e.g., time and cost budget). As EP8 noted: \textit{``I love that you have your guidelines from 60 people with disabilities. [Because having] a representative sample size is really tricky. I am an indie developer, and I only have two people who care about my game...It's important to talk to your community, get their input, and then develop based on what your actual customers' needs and desires are.''} With active involvement of the disability community, participants trusted the guidelines and were motivated to follow them. 

\textbf{\textit{Expanded User Bases.}}
Six participants (e.g., EP2, EP4, EP8) found that implementing the guidelines would attract a broader range of users. \change{User inclusion is particularly important for social VR platforms due to their core purpose of fostering socialization and connection}, as EP10 explained: \textit{``For a social VR platform, it is important to appeal to a wide audience by having that wide variety of avatars. If they just had five different avatars, they wouldn't be able to appeal to many people, and I think that they would not thrive. It is very important to have something where people can create an avatar that it either fulfills their player fantasy or make them feel like they're represented and included.''}
\change{Participants praised that our guidelines covered a variety of disabilities, including those that are commonly overlooked.} For example, EP1 was excited to see guidelines addressing the needs of people with mental health conditions (e.g., G2.2, G4.1). He commented on G4.1: \textit{``This guideline is really cool. I think it's especially helpful for people [with mental health conditions], since they've been traditionally underrepresented and are so hidden.''}

\textbf{\textit{Enhanced User Retention.}}
Participants believed that by following the guidelines, they could create more positive user experiences (EP1, EP2, EP4), contributing to the user retention rate \cite{ramli2022study}. Based on participants' professional experiences (EP1, EP2), users would develop a deeper attachment to their avatars when they well represented themselves. For example, EP1 pointed out the necessity of implementing G3.2 (i.e., allow detail customization of AT) to better engage the users: \textit{``I think if you are going to add assistive technology, you should allow a bit more customization over it to fit the particular personality or identity. [Because] it's going to make the users more engaged if they can have a more personalized avatar''}. 
As a result, participants were motivated to implement the guidelines to enhance their apps and promote user retention. 

\subsubsection{Determinants of Guidelines Implementation.} \label{determinants}
Although implementing guidelines introduced benefits, participants (e.g., EP2, EP4-6, EP8) had to consider \change{multiple} factors to decide the actual implementation practices and optimize the outcomes. As EP8 indicated: \textit{``As developers, we are limited to time, money, scope [of the applications], the platform limitations, and the hardware limitations, all of that.''} The implementation of guidelines is not an all-or-nothing choice. Instead, developers and designers may selectively follow guidelines that align with their goals and available resources. We reveal several factors that influence participants' decisions below.


\paragraph{\textbf{Alignment with Application Use Cases (D1)}}
Seven participants (e.g., EP5, EP6) indicated that guideline implementation should align with the application's use case, including its objectives (EP1, EP5) and design styles (EP3). For example, EP1 found that implementing G2.2 (i.e., enable expressive facial animations) was very important for apps that focused on VR therapy and mental health: \textit{``G2.2 is really relevant to my use case, where we're already showing internalized stigma in some patients' [avatar facial expressions]. The [avatar] may look like looking down, or being a little more hesitant to make eye contact when talking, or in general having muted body symptoms. I think this guideline is really important in that sense, because I've barely seen any of this being offered in some ways.''}
However, EP5 and EP7 believed G2.2 was only relevant to applications that focus on social communication. In other use cases like gaming, users may not pay attention to the avatar's facial expressions (EP5), making G2.2 a lower priority.

Moreover, three participants (EP1, EP3, EP5) mentioned that the guideline should match the overall \change{application theme and} artistic style. As EP1 commented on G1.4 (i.e., prioritize human avatars): \textit{``This guideline really depends on the use case. For some use cases, [using] human avatars might be totally illogical or contradictory to what [the application is] doing. For example, if it's a fantasy VR world about dragons, users might want to only have dragon avatars. Following the guideline of always [having a] human option could be overkill in some circumstances.''}

These examples suggest that guideline implementation is highly context-dependent. Participants considered a guideline important and actionable when it aligned well with both the objectives and \change{design style} of the application.

\paragraph{\textbf{Size of Affected User Groups (D2).}}
Although recognizing the importance of the guidelines, seven participants (e.g., EP6, EP7, EP9) emphasized the need to justify the use of guidelines based on the size of affected user groups. They reflected that many guidelines were \textit{``doable but just time consuming''} (EP9). Given the constraints on development resources, particularly time (EP9) and money (EP10), developers must ensure that the features they implement attract a viable number of users. As EP9 described: \textit{``These guidelines are important. They just need a justification for people to build them. Just being realistic, there have to be some reasons to put these many hours [to build a feature]. If you see your user demographics, and it's only one percentage of your audience, then [you would not want to spend that many hours].''} EP10 also indicated the limited budget in developing an application: \textit{``Games cost money to make, and they need to be able to appeal to a wide audience.''} When the development resources are limited, developers and designers would prioritize the guidelines that \change{impact a large amount of users}. 

\paragraph{\textbf{Noticeability of Features \change{(D3)}.}}
Four participants (EP1, EP2, EP3, EP4) prioritized implementing \change{guidelines that would deliver easy-to-notice features}. They see avatars as a medium to \change{attract attention} and stimulate social interactions. If other users cannot easily notice a feature on an avatar, they found no value in implementing it. For example, EP2 found the subtle behavior tracking in G2.1 was less important than the assistive technology design in G3.1: \textit{``If you are talking about slight movements like eye twitching, this is not something that is really noticeable. For example, if I have a prosthetic limb, I want to have it on my avatar. This is something that I can actually showcase, and [other people] will see... But if it is eye-twitching, it's not something that somebody is going to notice, and my importance rating is lower on things like that. Because in social VR, people go there to interact with other people and their environment. They are not just going to stare into the avatar's eyes or something else to see if it is twitching a little bit more or less.''}
As a result, participants generally preferred implementing the guidelines that suggested more visually dominant features (e.g., G3) than those with subtle features (e.g., G2.1, G2.2), unless the use cases required. 

\subsubsection{Suggested Improvements to the Guidelines.} \label{improvements}
While acknowledging the value of the guidelines and perceiving them as a great starting point, participants offered suggestions to improve the guidelines' clarity and actionability.

\paragraph{\textbf{\change{Provide a Concrete Scope} \change{(S1)}.}}
Eight participants (e.g., EP2, EP4, EP7) were concerned \change{about the wide coverage of certain guidelines (i.e., G1.3, G3.1, G3.2, G4.1) and the lack of concrete implementation scope. Such concerns primarily focused on guidelines that represent a broad range of disabilities, such as G3.1 (i.e., offer various types of assistive technology to cover a wide range of disabilities) and those promoting flexible customization, like G1.3 (i.e., enable customization of each body parts) and G3.2 (i.e., allow detailed customization of assistive technology). 
Participants were unsure about to what extent they should implement these guidelines (e.g., what assistive technology to include, how many customization options to provide).}
As EP4 explained: \textit{``There are so many disabilities, and they have so many different [variations]. It's very hard for designers to just cover all of them, or even most of them.''} 

Five participants (e.g., EP4, EP5, EP9) suggested to better specify the implementation scope for these broad guidelines. For instance, EP4 and EP5 desired G1.3 to have a checklist of body parts to customize and EP10 wanted a list of symbol examples to include for G4.1 (i.e., provide suitable icons, logos, and slogans that represent disability communities). 

However, two participants (EP1, EP5) pointed out a potential \change{dilemma between implementation feasibility and inclusion}, as specifying a certain scope may inadvertently marginalize the user groups that are not included in the list. As EP5 described: \textit{``It's really difficult to try to define that scope because there are so many disabilities. Trying to narrow them down to which are the best to represent, I feel that might even be more hurtful than helpful to some of the communities. But as a developer, if I wanted to try to actually take an action on some of these guidelines, I would need a defined scope that I should try to follow.''} One way to mitigate this dilemma, as suggested by EP1, is to use the scope as baseline \change{but encourage practitioners to include more options if time and resources allow}. 



\paragraph{\textbf{Diversify Implementation Examples \change{(S2)}.}}
Seven participants (e.g., EP4, EP8, EP10) found the guidelines on avatar body dynamics and assistive technologies (i.e., G2.2, G2.4, G3.3) 
to be less actionable. Specifically, EP5 and EP9 found that G2.2 (i.e., enable expressive facial animations) and G2.4 (i.e., leverage avatar posture to demonstrate the lived experiences) were largely restricted to the VR device's tracking capabilities. As EP5 reflected: \textit{``G2.4 does not infer any action needs to be done by the developer. Avatar motion is captured through motion tracking of the VR devices. What can or should be influenced to meet this guideline?''} When reflecting on the avatar facial expression for G2.2, EP10 discussed the technical challenges of inadequate tracking and rendering accuracy: \textit{``It's difficult to implement facial expressions that are accurate. Because the thing you need to do is the accurate [reflection] of feelings, but everybody's emotions changed dramatically and constantly. I just think with the current VR [equipment] that we have, it's going to be an extremely difficult thing to pull off emotions consistently.''}
Participants (EP1, EP2, EP4, EP5) thus recommended to diversify the avatar design examples for these guidelines to improve actionability. For example, EP5 would implement G2.2 by leveraging emotes and voice chat features to convey user status instead of relying on facial tracking. 

EP1 and EP5 also found that diversified examples can inspire them to generate more creative design ideas and learn more about the disability community.  
\change{For example, EP1 found G3.3 (i.e., provide high-quality simulation of assistive technologies to avoid misuse) could be supplemented with more examples to educate practitioners the stereotypical portrays: \textit{``Many developers might be new to this kind of inclusive design feature. They [probably don't know] if they're sufficiently following the guideline through. For example, I don't know all the stereotypes of different types of assistive technologies. As a developer, I might just happen to know one or two and then wouldn't know to research more. So showing more examples in addition to the wheelchair could be helpful.''}} 


\paragraph{\textbf{Add Applicable Use Cases \change{(S3)}.}} Beyond concrete avatar examples,  three participants (EP1, EP5, EP9) suggested to include the applicable use cases, guiding developers and designers to apply the guideline to suitable VR applications. As discussed in Section \ref{determinants}, use cases was a key factor influencing participants' implementation decision. EP1 found that clearly stating the suitable use cases for each guideline can enhance its applicability. 

\change{To better inform decision making for implementation, EP1 hoped the guidelines to suggest applicable social VR applications, allowing practitioners to prioritize the guidelines that fit for their use cases: \textit{``I [think the] way you would eventually frame these guidelines is to choose the ones that definitely apply to your applications. There might be a few guidelines that don't actually apply to your particular use case. For example, G1.4 (i.e., prioritizing human avatars) works well for certain kinds of games, especially for socially-oriented applications. But it's less applied to hardcore skill training.''}} 

\begin{table*}
\centering
\small
\begin{tabular}{|p{0.2cm}|p{0.5cm}|p{0.3cm}|p{6.3cm}|p{8.5cm}|}
\hline
\multicolumn{3}{|p{1.0cm}|}{} & \textbf{Design Guidelines} & \textbf{Description} \\
\hline

\multicolumn{5}{|>{\raggedright\arraybackslash}p{17.3cm}|}{\vspace{-2mm}\change{G0. Support Disability Representation in Social VR Avatars.
Approximately 1.3 billion people experience significant disability, representing about 16\% of the global population \cite{WHO2023}. It's important to ensure PWD are included and represented equally in emerging technology such as social VR. As long as the platform involves avatar-based interactions, there is design space to support disability representation.
}} \\ \hline

\multirow{4}{*}{\rotatebox[origin=c]{90}{\hspace{1em} \textbf{G1. Avatar Body Appearance} \hspace{1em}}} 


& G1.1
& \raisebox{0.2ex}{\change{HR}} 
& Default to full-body avatars to enable diverse disability representation across different body parts. 
&  The avatar interfaces should at least offer a full-body avatar option \cite{kelly2023}. We recommend making it the default or the starting avatar template, giving users the maximum flexibility to further customize their avatars as they prefer. 
\\ \cline{2-5}

& G1.2
& \raisebox{0.2ex}{\change{HR}}
& Enable flexible customization of body parts as opposed to using non-adjustable avatar templates. 
& Avatar interfaces should provide PWD sufficient flexibility to customize each avatar body part \cite{kelly2023}. \change{While the customization spans a wide range, the most commonly mentioned body parts to customize include (1) avatar height, (2) body shape, (3) limbs, and (4) facial features. 
} 
\\ \cline{2-5}

& G1.3
& \raisebox{0.2ex}{\change{HR}}
& Prioritize human avatars to emphasize the ``humanity'' rather than the ``disability'' aspect of identity. 
& Social VR applications \change{should offer human avatar options whenever the application theme allows.} 
\\ \cline{2-5}

& G1.4
& \raisebox{0.2ex}{\change{R}}
& Provide non-human avatar options to free users from social stigma in real life. 
& Besides human avatars, avatar interfaces should also provide diverse forms of non-human avatars, empowering PWD to choose the one they relate with flexibly. 
\\
\hline

\multirow{3}{*}{\rotatebox[origin=c]{90}{\hspace{0.1em} \textbf{G2. Dynamics} \hspace{1em}}} 
& G2.1
& \raisebox{0.2ex}{\change{HR}}
& Allow simulation or tracking of disability-related behaviors but only based on user preference.
& Users should be able to control the extent of behavior tracking in social VR. 
\\ \cline{2-5}

& G2.2
& \raisebox{0.2ex}{\change{R}}
& Enable expressive facial animations to deliver invisible status.
& Avatar platforms should enable diverse facial expressions, allowing PWD to express emotion, portray mental status, and indicate fluctuation of invisible disabilities \change{\cite{assets_24}}.
\\ \cline{2-5}

& G2.3
& \raisebox{0.2ex}{\change{HR}}
& Prioritize equitable capability and performance over authentic simulation. 
& 
Social VR platforms should ensure same level of capabilities and performance for all avatars no matter whether disability features or behaviors are involved.
\\ 
\hline

\multirow{5}{*}{\rotatebox[origin=c]{90}{\hspace{3em} \textbf{G3. Assistive Technology Design} \hspace{3em}}} 
& G3.1
& \raisebox{0.2ex}{\change{HR}}
& Offer various types of assistive technology to cover a wide range of disabilities.
& Avatar interfaces should offer assistive technologies that are commonly used by PWD \cite{zhang2022}. \change{The most desired types of assistive technologies should be included: (1) mobility aids; (2) prosthetic limbs; (3) visual aids; (4) hearing aids and cochlear implants; and (5) health monitoring devices.}
\\ \cline{2-5}

& G3.2
& \raisebox{0.2ex}{\change{HR}}
& Allow detail customization of assistive technology for personalized disability representation.
& Avatar platforms should allow customizations for assistive technology \change{\cite{kelly2023,zhang2022}. Basic customization options should include adjusting the colors of different assistive tech components and adding decorations to the assistive technologies} 
\\ \cline{2-5}

& G3.3
& \raisebox{0.2ex}{\change{HR}}
& Provide high-quality, authentic simulation of assistive technology to present disability respectfully and avoid misuse.
& To avoid misunderstandings or misuse, the assistive tech simulation should convey standardized, authentic details of the real-world assistive devices \change{\cite{zhang2023}, regardless the overall avatar style.} 
\\ \cline{2-5}

& G3.4
& \raisebox{0.2ex}{\change{R}}
& Demonstrate the liveliness of PWD through dynamic interactions with assistive technology.
& Beyond providing assistive tech options, social VR platforms should enable suitable interactions between avatar and the assistive technologies. The interactions should authentically reflect PWD's real-world usage of their assistive technologies.
\\ \cline{2-5}

& G3.5
& \raisebox{0.2ex}{\change{HR}}
& Avoid overshadowing the avatar body with assistive technology. 
& The size of assistive technology should not dominate the avatar body but rather fit the body size. 
\\
\hline

\multirow{5}{*}{\rotatebox[origin=c]{90}{\hspace{1em} \textbf{G4. Peripherals} \hspace{1em}}} 
& G4.1
& \raisebox{0.2ex}{\change{HR}}
& Provide suitable icons, logos, and slogans that represent disability communities. 
& Awareness-building items or presets should be provided, allowing users to attach them to various areas on or around the avatars. \change{A list of widely recognized and preferred symbols are: (1) the rainbow infinity symbol the represents the autism community \cite{assets_24, rainbow_infinity_symbol}, (2) the sunflower tha   t represents hidden disabilities \cite{isit_assets24, hidden_disability_sunflower}, (3) the disability pride flag \cite{disability_pride_flag}, (4) the spoons, symbolizing spoon theory for people with chronic illness \cite{kelly2023, assets_24}, and (5) the zebra symbols for rare diseases \cite{assets_24, Gualano_2023}.}
\\ \cline{2-5}

& G4.2
& \raisebox{0.2ex}{\change{R}}
& Leverage spaces beyond the avatar body to present disabilities.
& Developers and designers should consider leveraging avatar's peripheral space to enable users to better express their status, \change{especially for individuals with invisible disabilities.}
\\
\hline

\multirow{5}{*}{\rotatebox[origin=c]{90}{\hspace{1em} \textbf{G5. Interface} \hspace{1em}}} 
& G5.1
& \raisebox{0.2ex}{\change{HR}}
& Distribute disability features across the entire avatar interface rather than gathering them in a specialized category.
& Avatar features for disability expression should be treated in the same way to other avatar features. 
\\ \cline{2-5}

& G5.2
& \raisebox{0.2ex}{\change{HR}}
& Use continuous controls for high-granularity customization. 
&  Avatar interfaces should adopt input controls that offer a continuous range of options to enable flexible customization. 
\\ \cline{2-5}

& G5.3
& \raisebox{0.2ex}{\change{HR}}
& Offer an easy control to turn on/off or switch between disability features.
& Social VR platforms should provide easy-to-access shortcut control enable users to \change{conduct \textit{ad-hoc} avatar updates on the go.}
\\
\hline

\end{tabular}
\caption{\change{An overview of our \textit{revised} guidelines that incorporated VR practitioners' feedback. We present a set of 17 inclusive avatar guidelines for PWD that covered five avatar design aspects. Each guideline includes three elements: a recommendation level (Highly Recommended (HR) or Recommended (R)), guideline statement, and a detailed description. 
}}
\Description{The table is titled "An overview of our revised guidelines that incorporated VR practitioners' feedback." It presents a set of 17 inclusive avatar guidelines for PWD covering five avatar design aspects. Each guideline includes a recommendation level (Highly Recommended - HR, or Recommended - R), a guideline statement, and a detailed description.}
\label{tab:overview_revised}
\end{table*}


\paragraph{\textbf{Merging to Condense Guidelines (S4).}}
\change{Three participants (EP1, EP2, EP10) found three guidelines about assistive technologies (i.e., G3.1, G3.3, G3.4) and two guidelines about avatar dynamics (i.e., G2.1 and G2.4) contained redundant information that could be merged.
Specifically, EP2 found G3.3 (i.e., providing high-quality simulation of assistive technologies to avoid misuse) can be integrated into G3.1 (i.e., offer various types of assistive tech to cover a wide range of disabilities) by adding `high quality' suggestion to its description. While EP1 found both G3.3 and G3.4 (i.e., focus on simulating assistive technologies that empower PWD) discussed avoiding stereotypical portrays of PWD and thus could be combined into one guideline.

Moreover, two participants (EP1 and EP10) believed G2.4 (i.e., leverage avatar posture to demonstrate PWD's lived experiences) could be merged into G2.1 (i.e., allow simulation or tracking of disability-related behaviors based on user preferences), as they found posture was just a type of disability-related behaviors.}

\section{Guideline Revision}

\change{Based on the feedback from VR practitioners, we refined the guidelines and made the following changes: 
(1) Convert the original G1.1 to an overarching statement G0 to contextualize and motivate practitioners as well as specifying the application scope of our guidelines (D1);
(2) Assigned recommendation levels to each guideline to distinguish their priority based on development resources, size of user groups, and use cases (D1, D2, S3);
(3) Added statistics about served user groups, informing practitioners of the potential impact of implementing the guideline (D2);
(4) Specified the customization scope in G1.3, G3.1, G3.2, and G4.1 (S1); 
(5) Diversified implementation examples in G2.2, G2.3, and G3.3 (S2);
(6) Suggested potential use cases for G1.4 and G2.2 (D1, S3);
(7) Merged G2.4 into G2.1 to remove redundant information, and merged G3.1 and G3.4 to clarify the characteristics of simulated assistive technologies in social VR (S4);
\cameraready{(8) Refined the wording in G2.2, G3.3, and G5.2 to improve clarity.
Appendix Table \ref{tab:changes} demonstrated the changes we made to the initial guidelines.}

After revision, we ended up with 17 design guidelines upon experts' feedback. We present an overview table of the revised guidelines in Table ~\ref{tab:overview_revised} and a full version in Appendix Table ~\ref{tab:full_revised}.}

\section{Discussion}

\change{We derived a centralized, comprehensive, and validated set of guidelines with concrete evidence and actionable implementation examples. Through a systematic literature review and interview with 60 PWD,}
our guidelines thoroughly covered diverse disability expression strategies through five design aspects, including avatar appearance (G1), body dynamics (G2), assistive technology design (G3), peripherals around avatars (G4), and avatar interface design and customization controls (G5). 
Moreover, we evaluated and iterated the guidelines with 10 AR/VR experts. The evaluation results suggested that: (1) the guidelines comprehensively covered a broad range of disability expression methods through avatar designs; (2) practitioners in the industry can use guidelines to identify real problems in existing avatar platforms, proving that the guidelines are applicable; (3) the guidelines were easy to understand and actionable for avatar development in practice. 

\change{Based on experts' feedback, we further revised the guidelines to make them more concise and actionable for industry practitioners. The finalized version contains 17 revised design guidelines with detailed descriptions, concrete evidence, recommendation levels, and implementation examples.} We believe the guidelines can serve as a valuable resource that supports industry practitioners in designing inclusive avatars for PWD. 

In the following sections, we discuss the guidelines that spurred interesting conversations among practitioners, delving into their implementation considerations and  applicable use cases, and envisioning future directions.

\change{\textbf{Centralized Guidelines with Actionable Implementation Details.}
Building upon the valuable design implications derived by prior works \cite{zhang2022, zhang2023, kelly2023, assets_24}, we further advance the field of inclusive avatar designs by 1) centralizing relevant design implications, 2) strengthening them with implementation details, and 3) adding new design guidelines.

When generating the guidelines, we cross-referenced the prior implications with our interview findings, aiming for a comprehensive and centralized set of guidelines that are easy to apply, share, and update. Such centralized guidelines are pressingly needed for industry practitioners, as they often have limited access to, and find it hard to locate such resources, which are the two main challenges in translating academic research into industry practices~\cite{Colusso2017}. To better integrate inclusion in VR industry, more translational efforts are needed in future research, such as website collecting various resources \cite{xraccess} and open-source toolkits \cite{zhao2019seeingvr}, to enable easy access to high-quality resources for the practitioners. 

Among the finalized 17 guidelines, seven (G1.3, G1.4, G2.1, G2.3, G3.4, G5.1, G5.2) 
are newly generated from interview findings, and 11 are expanded from prior literature with concrete evidence (e.g., applicable use cases, served user groups), scope, and implementation examples. For example, both Zhang et al. \cite{zhang2022} and Mack et al. \cite{kelly2023} suggested offering diverse types of assistive technologies (G3.1). We concretize this implication by specifying a list of the five most preferred types of assistive technologies, making it more actionable with a concrete implementation scope. 
With the collective effort of prior works and our supplemental large-scale interview, we believe this set of guidelines can serve as a valuable resource for industry practitioners to practice upon.}

\change{\textbf{Flexible Applications based on Use Cases.}} Our guidelines aim to optimize for the preferred identity representations of PWD in social VR contexts, and we do not expect all platforms to implement every guideline. We \change{recognize the compounding factors (e.g., cost-effectiveness, development effort, user size) that may affect practitioners' decisions of following a guideline.} We also anticipate situations where the suggestions in the guidelines may not align well with the overall theme or aesthetic style of a social VR platform. For example, prioritizing human avatars as suggested in G1.4 may conflict with platforms featuring fantastical themes. In these cases, developers and designers may choose to follow the guidelines that apply to their use cases. \change{In the revised guidelines, we attempted to resolve this tension by assigning recommendation levels (i.e., highly recommended and recommended) to each guideline, clarifying their priority to better inform practitioners' decision-making.}

\change{\textbf{Balance Safety and Diverse Representation}.} Our guidelines briefly touch on the safety issues in expressing disabilities on avatars. The avatar-based disability representation could trigger explicit and embodied harassment targeting at PWD \cite{zhang2023}. Our guidelines proposed some reactive ways to mitigate the negative experiences, such as G5.3 (i.e., offering easy control to turn on and off disability features). 
Future work should explore more proactive strategies to prevent harassment from happening when using avatars with disability-related features. It is imperative to ensure PWD represent themselves safely in social VR.

\change{\textbf{Multimodal Disability Representations.}} Lastly, we note that our work primarily focus on deriving guidelines through the visual aspects of avatar designs. Beyond visual features, avatar voice is another critical channel that can reveal and represent a user's identity in real life \cite{Povinelli_voice_2024}. \change{However, using voice for disability representation could lead to possible tensions and conflicts. For example, people with speech disabilities (e.g., stuttering) may not want to disclose their disabilities, which often comes with substantial social penalties in real life, such as negative listener reactions and teasing \cite{wu2023stuttering}. Moreover, avatar voice could unintentionally reveal a person's voluntary identities, leading to negative user experiences \cite{Maloney2020Anonymity, freeman2022disturbing}. In these cases, some of our guidelines may not be fully applicable, such as G2.1 (i.e., allow simulation or tracking of disability-related behaviors based on user preferences).} 
We encourage future work to further study identity construction through voice in social VR. We hope the methodologies and outcomes presented in this work could inform and inspire future research to develop guidelines for inclusive social VR through multi-modal perspectives.









\section{Conclusion}

We presented a centralized, comprehensive, and validated set of 17 design guidelines for inclusive avatar design. Through a systematic literature review and interview with 60 PWD, the guidelines covered comprehensive disability expression methods through five design aspects of social VR avatars. With the emerging XR technology, we see the guidelines' potentials in shaping design standards and providing implementation insights for a more inclusive and diverse social VR. 

\begin{acks}
We thank all anonymous participants for their efforts and valuable feedback. This work was supported in part by the National Science Foundation under Grant No. IIS-2328182 \& No. IIS-2328183 and a Meta Research Award.
\end{acks}

\bibliographystyle{ACM-Reference-Format}
\bibliography{reference}


\appendix
\section{Appendix}
 \subsection{Study I: Interview with PWD} \label{protocol_pwd}
\subsubsection{Introduction Phase}
\begin{enumerate}
    \item What’s your age?
    \item How do you identify your gender? 
    \item What’s your ethnicity? 
    \item What disability do you have?
    \item Do you have any episodic disabilities, where the type or severity of symptoms fluctuate and cause a variation of abilities?
        \begin{itemize}
            \item If so, in what ways?
        \end{itemize}
    \item What social VR platforms have you used?
    \item In general, what aspects of your identities would you like to convey through avatar? (prompt: Do you want to reveal gender through avatars? Disability? Race? Age?)
        \begin{itemize}
            \item (Go through each of the mentioned identities:) Why? 
            \item How do you want to represent the disability? 
            \item (for disability, if not:) why not?
        \end{itemize}
    
\end{enumerate}

\subsubsection{Avatar Design Phase}
In the following, we will focus on understanding your preferences for presenting disability identities on avatars. We will delve into each aspect of avatars and ask how you would like to design your ideal avatars. We will go through from a higher level of general avatar types to the details of each component of an avatar. Please think through with me, imagine me as a design agent, and we will discuss the avatar aspects step by step. Let’s start with the general types of avatars.

\textbf{Body Appearance:}
\begin{enumerate}
    \item What types of avatar do you want?
        \begin{itemize}
            \item Why?
            \item How does it relate to your disability identity representation, if at all? 
        \end{itemize}
    \item What body parts do you want your avatar to have? 
        \begin{itemize}
            \item Why?
            \item How does this physical composition of your avatar reflect your disability, if at all?
        \end{itemize}
    \item What’s the desired skin color of your avatar? 
        \begin{itemize}
            \item Does the skin color relate to your disability identity or its intersection with other identities? If so, how do you think it could relate? 
        \end{itemize}
    \item How tall do you want your avatar to be? 
        \begin{itemize}
            \item Why? 
            \item How does avatar's height reflect your disability, if at all?
            \item What’s the preferred body shape of your avatar?
            \item What are your preferences for the size and proportion of your avatar’s different body parts?
                \begin{itemize}
                    \item Do you have any body parts that you want to highlight/distinguish in comparison to other parts? OR: Do you want to customize any body parts? 
                    \item Why?
                \end{itemize}
            \item How does the choice of body shape, body size, and body proportion reflect your disability, if at all?
        \end{itemize}
    \item Considering the facial features of your avatar, how do you want to design the eyes? How about cheekbones, nose, mouth-lips, forehead, jawline of your avatar? (go through each of them with the following question; if not, just skipped)
        \begin{itemize}
            \item Why? 
            \item How does the appearance of eyes, cheekbones, mouth, forehead, and jaw reflect your disability, if at all? 
        \end{itemize}
    \item What’s the preferred hairstyle for your avatar?
        \begin{itemize}
            \item How does the hairstyle reflect your disability, if at all?
        \end{itemize}
\end{enumerate}

\textbf{Body Motion:}
\begin{enumerate}
    \item What specific movements and expressions would you like to see in the avatar's facial features, such as the eyes, mouth, and eyebrows? 
        \begin{itemize}
            \item How do the facial animations of the avatar reflect your disability, if at all?
        \end{itemize}
    \item What do you want your avatar's movement to be like? 
        \begin{itemize}
            \item How do you feel about the movement of the avatar's limbs? (prompt: ease of movement? The range of motion?)
            \item How does the movement of avatars reflect your disability, if at all?
        \end{itemize}
\end{enumerate}

\textbf{Assistive Technology:}
\begin{enumerate}
    \item Do you use any assistive technologies (AT) in real life?
    \item Do you want to add any AT to avatars?
        \begin{itemize}
            \item If yes, what are the AT you want to add?
            \item Why do you want to add the AT?
            \item How would you like the AT to look like? 
                \begin{itemize}
                    \item What do you think would be the proper size of the AT to your avatar body? What’s the ratio?
                    \item Do you have any specific aesthetic preferences for the AT? (e.g., color, brand)
                    \item How would you like the AT to be positioned?
                    \item Do you expect the AT to be interactive?
                        \begin{itemize}
                            \item If so, how do you imagine to interact with it?
                            \item Do you want others to interact with your AT? If so, how do you expect other people to interact with it? 
                        \end{itemize}
                \end{itemize}
                \begin{itemize}
                    \item If not, why?
                \end{itemize}
        \end{itemize}
    \item Do you have any other ideas about the AT design that you would like to share? 
\end{enumerate}

\textbf{Outfit, Accessories, and Avatar Surroundings:}
\begin{enumerate}
    \item How would you like to design the outfit of avatars? 
        \begin{itemize}
            \item Why do you want to design the outfit in this way? 
            \item Do you think the outfit design could reflect your disability? 
            \item If so, how? 
        \end{itemize}
    \item Would you like to have any accessories on avatars?
        \begin{itemize}
            \item If so, what are they? 
            \item Why do you want to have these accessories? 
            \item How do they represent your disability, if at all?
        \end{itemize}
    \item Do you think anything surrounding the avatar, instead of directly on/attached to the avatar, would help represent your disabilities? 
        \begin{itemize}
            \item If so, how? What do they look like?
        \end{itemize}
    \item Is there anything else you would like to have on your avatar for disability representation?
\end{enumerate}

\textbf{Multiple and Intersectional Identities:}
\begin{enumerate}
    \item If your disability condition fluctuates, how do you want to present the fluctuating conditions via avatar design, if at all?
    \item If you have multiple disabilities, how would you like to present them, if at all? 
    \item As each person’s identity is multifaceted (such as gender, race, disability, etc.), have you encountered or do you envision encountering any conflicts in representing multiple identities via avatar design?
        \begin{itemize}
            \item If so, what are the conflicts? 
            \item How do you deal with them through avatars?
            \item What features do you think can mitigate such conflicts?
        \end{itemize}
\end{enumerate}

\subsubsection{Reflection Phase}
\begin{enumerate}
    \item What factors would affect your willingness/decision to disclose disability via avatars?
    \item How do you want to interact with the avatar customization interface? 
    \item Are there any other design aspects that you think can represent your disabilities that we haven’t mentioned? 
    \item Do you have any other suggestions about how to design inclusive avatar for people with disabilities? (Prompt: What should we do? What should we not do?)
\end{enumerate}

\subsection{Study II: Protocol with VR Experts}

\subsubsection{Heuristic Evaluation Survey}
\label{heuristic_survey}
[Survey page starts with the presentation of a guideline.]
\begin{enumerate}
    \item How do you interpret this guideline?
    \item Do you have any questions or anything you found confusing about this guideline? Please share your thoughts below.
    \item Does this guideline apply (e.g., is it relevant) to the social VR avatars you reviewed? 
        \begin{itemize}
            \item Yes
            \item No (i.e., irrelevant or our of scope)
        \end{itemize}
    \item If not, can you explain why do you think this guideline does not apply to the avatars you reviewed? 
    \item Based on the social VR app you reviewed, how well does its avatars apply (i.e., follow or implement) this guideline? Please rate it on a scale of 1 to 5. 


    \item Please provide an example to describe how it applies. (We encourage you to revisit the application and take screenshots to demonstrate the example. Please also feel free to annotate on the screenshots.)
    \item Please provide an example to describe how it violates. (We encourage you to revisit the application and take screenshots to demonstrate the example. Please also feel free to annotate on the screenshots.)
    \item As a developer/designer, please rate the following aspects of this guideline based on your experiences in developing and designing avatars for AR/VR applications. Please use the scale from 1 (worst) to 5 (best) for each dimension: 
        \begin{itemize}
            \item Level of Clarity - How clear and understandable is the guideline? (1 = very confusing, and 5 = very clear)
            \item Level of Actionability - How actionable do you find the guideline? (1 = very unactionable, and 5 = very actionable)
            \item Level of Importance - How important do you think it is to follow this guideline? (1 = very unimportant, and 5 = very important)
        \end{itemize}


%

    \item Could you explain your rating of the clarity, actionability, and importance of this guideline?

\end{enumerate}

\subsubsection{Interview Protocol} \label{protocol_experts}

\begin{enumerate}
    \item How old are you? 
    \item What’s your ethnicity?
    \item In total, how long have you worked as a full-time AR/VR developer/designer?
    \item Is your current job centered on AR/VR development?
        \begin{itemize}
            \item If yes, 
            \begin{itemize}
                \item What company do you work for?
                \item What is your position in this organization?
                \item Can you briefly describe what you do for this position?
                \item How long have you been working at this company?
            \end{itemize}
            \item If not, 
            \begin{itemize}
                \item What’s your most recent AR/VR related job?
                \item What company did you work for?
                \item What was your position?
                \item How long did you work at this company? 
            \end{itemize}
        \end{itemize}
    \item Do you have any experience with avatar modeling, animation, or design? 
        \begin{itemize}
            \item If yes, could you briefly describe your experience?
            \item What development platforms have you used?
            \item Have you ever designed avatars for people with disabilities? or for any other underrepresented groups? Can you briefly describe it if so?
        \end{itemize}
\end{enumerate}

\textbf{Guidelines Evaluation and Survey Follow-up}
[Ask the following questions for each guideline:]
\begin{enumerate}
    \item In [social vr app], do you think its avatar applies (or follows) this guideline? 
        \begin{itemize}
            \item If yes: could you revisit [social vr app] and show us an example of how it applies? 
            \item If not:  could you explain your answer? 
        \end{itemize}
    \item In [social vr app], do you think its avatar violates (or breaks) this guideline?
        \begin{itemize}
            \item If yes: could you revisit [social vr app] and show us an example of how it violates?
            \item If not: could you explain your answer? 
        \end{itemize}
    \item How clear and understandable is this guideline? Please rate it on a scale of 1 to 5, with 1 being very unclear and 5 being very clear. 
        \begin{itemize}
            \item Could you explain your choice?
            \item If unclear, what confused you specifically? [prob: wording/phrasing? Or need further explanation?] 
            \item How can we improve this guideline to make it more clear? 
        \end{itemize}
    \item As a developer/designer, how actionable do you find this guideline is? Please rate it on a scale of 1 to 5, with 1 being very not actionable and 5 being very actionable. 
        \begin{itemize}
            \item Can you explain your choice? 
            \item If actionable, can you describe how you want to apply this guideline to the avatar systems in practice?
            \item If not actionable, what information do you need to implement this guideline?
        \end{itemize}
    \item How important do you think it is to follow this guideline when you develop/design avatars for your AR/VR applications? Please rate it on a scale of 1 to 5, with 1 being very unimportant and 5 being very important.
        \begin{itemize}
            \item Can you explain your choice? Why do you think it’s important/not important?
        \end{itemize}
\end{enumerate}

\textbf{Exit Interview}

\begin{enumerate}
    \item Is there anything that you think is important but we haven’t included here?
        \begin{itemize}
            \item If so, what are they?
        \end{itemize}
    \item Have you noticed any redundancies in the guidelines?
        \begin{itemize}
            \item If yes, what are they? Which parts make you feel they are redundant? 
            \item Why do you think they are redundant?
            \item How would you like to improve them?
        \end{itemize}
    \item Have you noticed any guidelines conflicting with each other? 
        \begin{itemize}
            \item If so, what are they? 
            \item Why do you think they are conflicting with each other?
            \item How would you like to improve them?
        \end{itemize}
    \item In general, do you think the language (wording and phrasing) is easy to understand? 
        \begin{itemize}
            \item Any suggestions to further improve the language?
        \end{itemize}
    \item Currently each guideline has four components: guideline-explanation-examples-visual demos. What do you think of this guideline structure?
        \begin{itemize}
            \item How would you like to further improve the structure? (prob: add/remove any components?)
        \end{itemize}
    \item Anything you like about this guideline?
    \item How should we further improve the guidelines to make it easier for you to understand and implement it in your project? 
\end{enumerate}

\subsection{Tables and Figures}

\newpage
\begin{table}[h!]
\centering
\small
\onecolumn
\begin{tabular}{p{2cm}p{2.3cm}p{2.1cm}p{5cm}p{4.5cm}}
\toprule
\textbf{Social VR Apps} & \textbf{Avatar Types} & \textbf{Components} & \textbf{Customization Options} & \textbf{Disability Representation} \\
\toprule
VRChat & Human/animals/ furry/robotic avatar, etc.  & Full body & Selection from default avatar interface; self-upload avatars & No default disability representations; allow self-upload avatars with disability features \\
\hline
Horizon Worlds & Cartoonish human avatar & Full body & Selection from default avatar interface; offer body appearance and facial details customization & Cochlear implants and hearing aids with color customization \\
\hline
Rec Room & Cartoonish human avatar & Upper body with floating hands & Selection from default avatar interface; offer body appearance and facial details customization with paywall & No default disability representations \\
\hline
Roblox & Lego-character-like avatar  /animal avatar & Full body & Selection from default avatar interface; offer body appearance and facial details customization with paywall & Wheelchair options; crutches; emotes with diverse emotions expressions; animation bundles \\
\hline
Multiverse & Robotic avatar & Upper body with floating hands & No avatar customization & No disability representation \\
\bottomrule
\end{tabular}
\Description{The table titled "Table 4. Review of Selected Social VR Platforms and Avatar Features" provides a comparative analysis of various social VR applications, focusing on avatar types, customization options, components, and disability representation.}
\caption{Review of Selected Social VR Apps and Avatar Features. Five mainstream social VR apps are reviewed for their avatar types, components, customization options, and offered disability representations.}
\label{tab:vr_platforms}
\end{table}

\begin{table*}[!t]
\centering
\onecolumn
\small 
\begin{tabular}{|p{0.5cm}|p{0.4cm}|p{6.8cm}|p{8.5cm}|}
\hline
\textbf{} & \textbf{} & \textbf{Design Guidelines} & \textbf{Examples} \\
\hline
\multirow{5}{*}{\rotatebox[origin=c]{90}{\hspace{1em} \textbf{G1. Body Appearance} \hspace{1em}}} 
& G1.1
& Support disability representation in social VR avatars.
& Avatar system provides hearing devices for disability representation. \\ \cline{2-4}

& G1.2
&  Default to full-body avatars to enable diverse disability representation across different body parts.
& A full-body avatar can show a prosthetic left foreleg.  \\ \cline{2-4}

& G1.3
& Enable flexible customization of body parts as opposed to using non-adjustable avatar templates. 
& Avatar platforms should offer options to customize the size of each eye. \\ \cline{2-4}

& G1.4
& Prioritize human avatars to emphasize the ``humanity'' rather than the ``disability'' aspect of identity.
& VALID validated avatar library (left) and Ready Player Me avatar (right) present human avatars that can show intersecting identities of disabilities, race, and gender.  \\ \cline{2-4}

& G1.5
& Provide non-human avatar options to free users from social stigma in real life 
& A robotic avatar can represent left forearm amputation.  \\
\hline

\multirow{4}{*}{\rotatebox[origin=c]{90}{\hspace{0.1em} \textbf{G2. Avatar Dynamics} \hspace{0.1em}}} 
& G2.1
& Allow simulation or tracking of disability-related behaviors but only based on user preference.
& Avatar can show motor tics or not based on the user's preference. \\ \cline{2-4}

& G2.2
& Enable expressive facial animations that deliver a spectrum of emotions.
& Avatar can show diverse emotions, including anxiety, sadness, and happiness. \\ \cline{2-4}

& G2.3
& Prioritize equitable capability and performance over authentic simulation.
& The avatar with the wheelchair can move at the same speed as the avatar without the wheelchair.\\ \cline{2-4}

& G2.4
& Leverage avatar posture and motion to demonstrate the lived experiences of people with disabilities.
& The avatar representing a low vision person can show a different posture than the avatar representing a sighted person in a conversation.   \\
\hline

\multirow{6}{*}{\rotatebox[origin=c]{90}{\hspace{1em} \textbf{G3. Assistive Technology Design} \hspace{1em}}} 
& G3.1
& Offer various types of assistive technology to cover a wide range of disabilities.
& Users can add multiple types of mobility aids, such as crutches and prosthetic limb, to their avatars. \\ \cline{2-4}

& G3.2
& Allow detail customization of assistive technology for personalized disability representation.
& Users can adjust the color of the power wheelchair, like the cushions, wheels, and chassis cover. \\ \cline{2-4}

& G3.3
& Provide high-quality, realistic simulation of assistive technology to present disability respectfully and avoid misuse.
& The design of a white cane for people with low vision should follow the standardized color selection for such walking aids.\\ \cline{2-4}

& G3.4
& Focus on simulating assistive technologies that empower people with disabilities, rather than those that highlight their challenges.
& The manual wheelchair should not have handles, demonstrating that it’s designed for users who want to navigate independently instead of being pushed. \\ \cline{2-4}

& G3.5
& Demonstrate the liveliness of PWD through dynamic interactions with assistive technology.
& Avatar controls the manual wheelchair through pushing. \\ \cline{2-4}

& G3.6
& Avoid overshadowing the avatar body with the assistive technology. 
& Users can change the size of assistive technology to match with their avatar size.  \\
\hline

\multirow{2}{*}{\rotatebox[origin=c]{90}{\parbox{1.4cm}{\centering \textbf{G4. Peri-}\\\textbf{pherals}}}} 
& G4.1
& Provide suitable icons, logos, and slogans that represent disability communities.
& An avatar wearing a T-shirt with a rainbow and infinity symbol to represent the autism spectrum disorder community. \\ \cline{2-4}

& G4.2
& Leverage spaces beyond the avatar body to present disabilities.
& An avatar with a floating bubble overhead, showing a level of social energy. \\
\hline

\multirow{3}{*}{\rotatebox[origin=c]{90}{\parbox{2.5cm}{\centering \textbf{G5. Interface}}}} 
& G5.1
& Distribute disability features across the entire avatar interface rather than gathering them in a specialized category.
& Walking canes and wheelchairs are included under the accessory category along with items like glasses, hats and bags. \\ \cline{2-4}

& G5.2
& Use input controls that offer precise adjustments whenever possible.
& Avatar should offer a slider to change the length of limbs. \\ \cline{2-4}

& G5.3
& Offer an easy control to turn on/off or switch between disability features.
& Users should be able to turn disability-related features on and off with a single click. \\
\hline

\end{tabular}
\caption{Our \change{\textit{initial}} 20 design guidelines for inclusive avatars. }
\Description{The table titled "Table 5. Our original 20 design guidelines for inclusive avatars" presents 20 guidelines grouped into five categories: Avatar Body Appearance, Avatar Dynamics, Assistive Technology Design, Peripherals, and Interface. Each guideline labeled with identifiers (e.g., G1.1, G2.2). The "Examples" column illustrates each guideline with practical implementations, such as using full-body avatars to represent prosthetic limbs or allowing customization of assistive technology components. Each row corresponds to a specific guideline, presenting a structured layout that aligns guidelines with illustrative use cases for creating inclusive avatars.}
\label{tab:overview_original}
\end{table*}

\begin{figure}[!ht]
    \centering
    \begin{subfigure}[b]{0.495\textwidth}
        \centering
        \includegraphics[width=\textwidth]{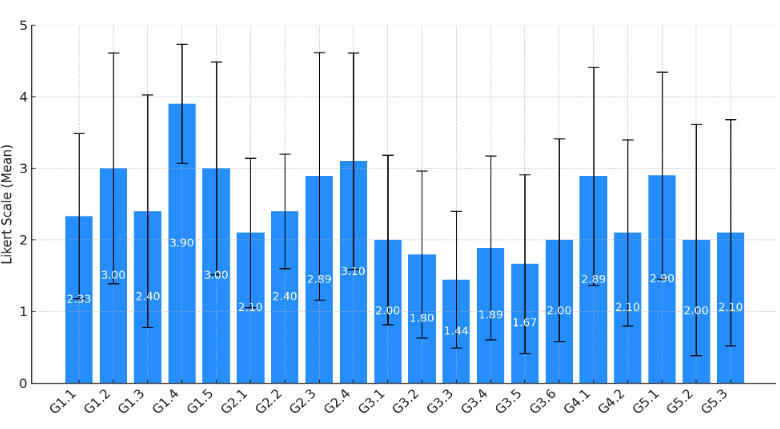}
        \caption{Level of Application}
        \label{fig:application}
    \end{subfigure}
    \hfill 
    \begin{subfigure}[b]{0.495\textwidth}
        \centering
        \includegraphics[width=\textwidth]{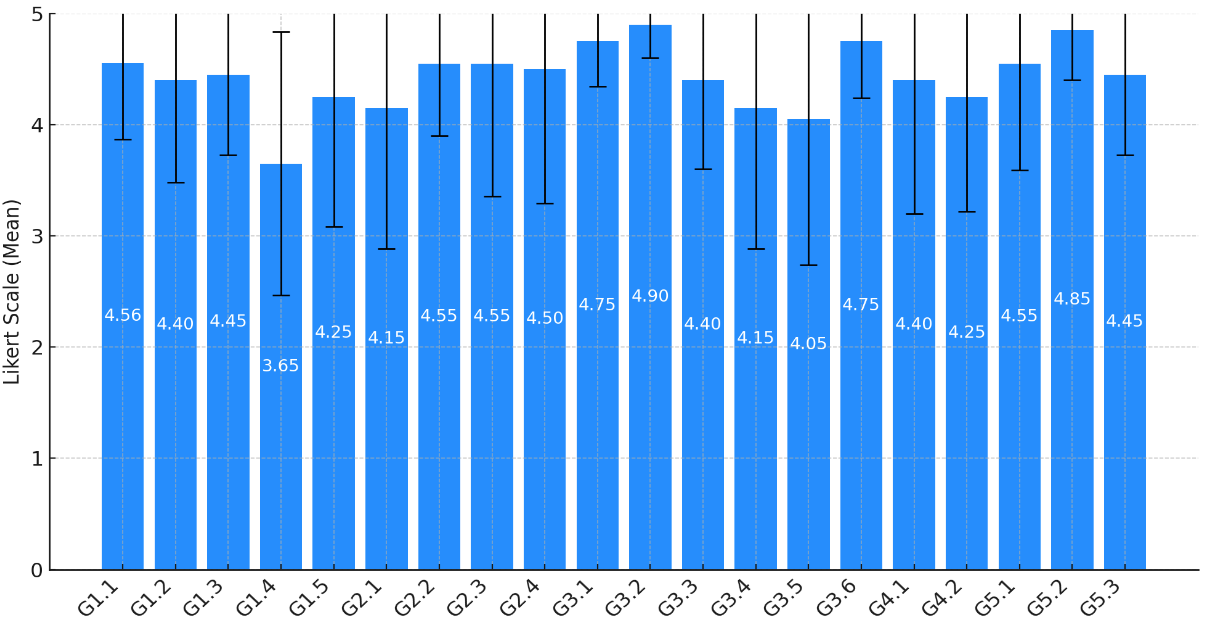}
        \caption{Level of Clarity}
        \label{fig:clarity}
    \end{subfigure}
    \newline 
    \begin{subfigure}[b]{0.495\textwidth}
        \centering
        \includegraphics[width=\textwidth]{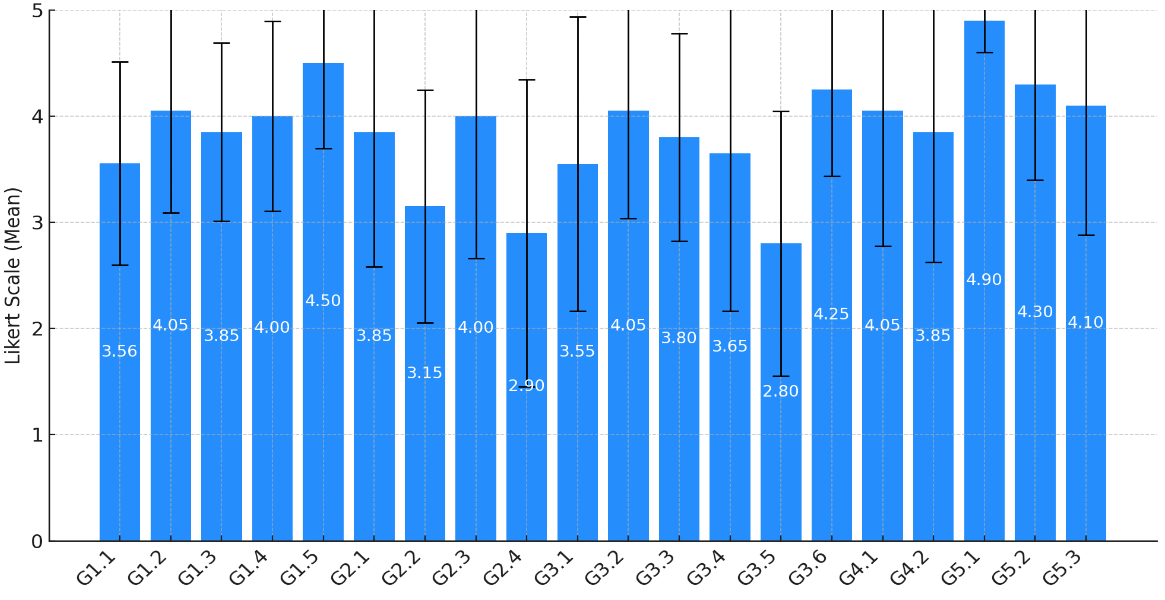}
        \caption{Level of Actionability}
        \label{fig:actionability}
    \end{subfigure}
    \hfill
    \begin{subfigure}[b]{0.495\textwidth}
        \centering
        \includegraphics[width=\textwidth]{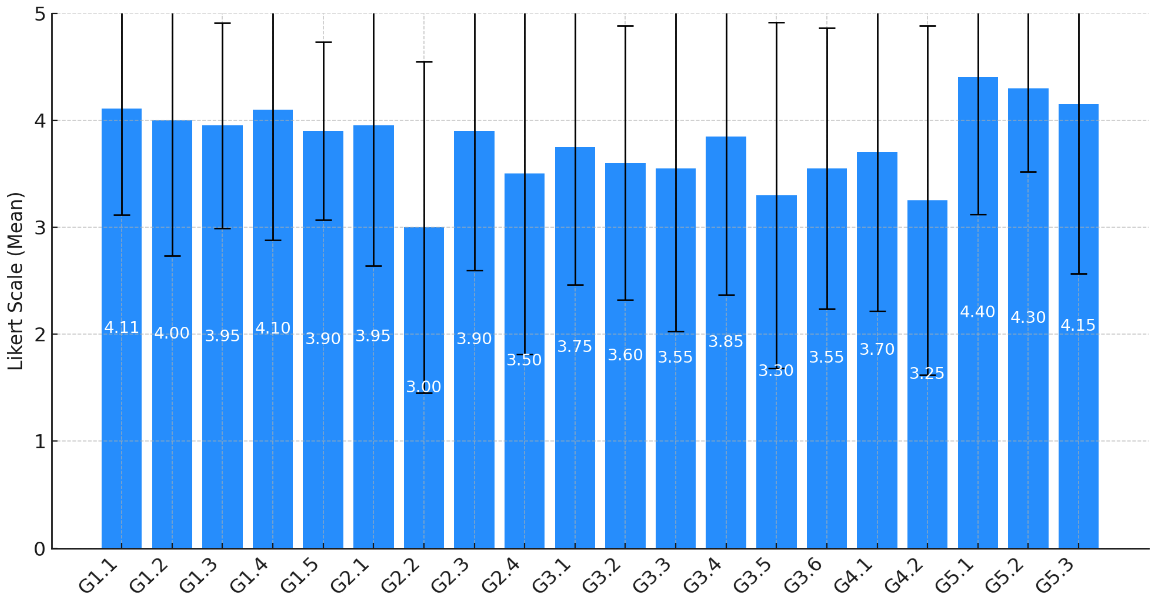}
        \caption{Level of Importance}
        \label{fig:importance}
    \end{subfigure}
    \caption{\change{Mean rating of applications (top left), clarity (top right), actionability (bottom left), and importance (bottom right) for each guideline. X-axis shows the ID of each guideline, and y-axis shows the mean value of 5-point Likert scale.}}
    \Description{This figure presents four bar charts evaluating various design guidelines based on four metrics: Level of Application, Level of Clarity, Level of Actionability, and Level of Importance. Each chart uses a Likert scale (mean values) with error bars representing variability.}
    \label{fig:complete}
\end{figure}

\begin{table*}
\centering
\onecolumn
\small 
\begin{tabular}{|p{0.5cm}|p{5.9cm}|p{5.9cm}|p{3.9cm}|}
\hline
\textbf{} & \textbf{Initial Guidelines} & \textbf{Revised Guidelines} &\textbf{Changes}\\
\hline
\multirow{5}{*}{\rotatebox[origin=c]{90}{\hspace{1em} \textbf{G1. Body Appearance} \hspace{1em}}} 
& G1.1. Support disability representation in social VR avatars.
& G0. Support disability representation in social VR avatars. 
& Changed the original G1.1 to an overarching statement G0 to contextualize and motivate practitioners as well as specifying the application scope of the guidelines. \\ \cline{2-4}

& G1.2. Default to full-body avatars to enable diverse disability representation across different body parts. 
 & G1.1. Default to full-body avatars to enable diverse disability representation across different body parts. 
&  \\ \cline{2-4}

& G1.3. Enable flexible customization of body parts as opposed to using non-adjustable avatar templates.
& G1.2. Enable flexible customization of body parts as opposed to using non-adjustable avatar templates.
&  \\ \cline{2-4}

& G1.4. Prioritize human avatars to emphasize the ``humanity'' rather than the ``disability'' aspect of identity.
& G1.3. Prioritize human avatars to emphasize the ``humanity'' rather than the ``disability'' aspect of identity.
&  \\ \cline{2-4}

& G1.5. Provide non-human avatar options to free users from social stigma in real life.
& G1.4. Provide non-human avatar options to free users from social stigma in real life.
&  \\
\hline

\multirow{4}{*}{\rotatebox[origin=c]{90}{\hspace{0.5em} \textbf{G2. Avatar Dynamics} \hspace{0.5em}}} 
& G2.1. Allow simulation or tracking of disability-related behaviors but only based on user preference.
& G2.1. Allow simulation or tracking of disability-related behaviors but only based on user preference.
&  \\ \cline{2-4}

& G2.2. Enable expressive facial animations that deliver a spectrum of emotions.
& G2.2. Enable expressive facial animations to deliver invisible status.
& Changed the phrase ``a spectrum of emotions'' to ``invisible status'' to describe PWD's preferences more accurately. \\ \cline{2-4}

& G2.3. Prioritize equitable capability and performance over authentic simulation.
& G2.3. Prioritize equitable capability and performance over authentic simulation. 
&  \\ \cline{2-4}

& G2.4. Leverage avatar posture and motion to demonstrate the lived experiences of PWD.
& \st{G2.4. Leverage avatar posture and motion to demonstrate the lived experiences of PWD.}
& Merged the original G2.4 into G2.1. to remove redundant information. \\
\hline

\multirow{6}{*}{\rotatebox[origin=c]{90}{\hspace{1em} \textbf{G3. Assistive Technology Design} \hspace{1em}}} 
& G3.1. Offer various types of assistive technology to cover a wide range of disabilities.
& G3.1. Offer various types of assistive technology to cover a wide range of disabilities.
&  \\ \cline{2-4}

& G3.2. Allow detail customization of assistive technology for personalized disability representation.
& G3.2. Allow detail customization of assistive technology for personalized disability representation.
&  \\ \cline{2-4}

& G3.3. Provide high-quality, realistic simulation of assistive technology to present disability respectfully and avoid misuse.
& G3.3. Provide high-quality, authentic simulation of assistive technology to present disability respectfully and avoid misuse.
& Changed the word ``realistic'' to ``authentic'' to better describe the preferred characteristics of assistive technology designs. \\ \cline{2-4}

& G3.4. Focus on simulating assistive technologies that empower PWD, rather than those that highlight their challenges.
& \st{G3.4. Focus on simulating assistive technologies that empower PWD, rather than those that highlight their challenges.}
& Merged the original G3.4 to G3.1 to remove redundant information. \\ \cline{2-4}

& G3.5. Demonstrate the liveliness of PWD through dynamic interactions with assistive technology. 
& G3.4. Demonstrate the liveliness of PWD through dynamic interactions with assistive technology.
& 
\\ \cline{2-4}

& G3.6. Avoid overshadowing the avatar body with the assistive technology. 
& G3.5. Avoid overshadowing the avatar body with the assistive technology.
&  \\
\hline

\multirow{2}{*}{\rotatebox[origin=c]{90}{\parbox{1.4cm}{\centering \textbf{G4. Peri-}\\\textbf{pherals}}}} 
& G4.1. Provide suitable icons, logos, and slogans that represent disability communities. 
& G4.1. Provide suitable icons, logos, and slogans that represent disability communities.
&  \\ \cline{2-4}

& G4.2. Leverage spaces beyond the avatar body to present disabilities.
& G4.2. Leverage spaces beyond the avatar body to present disabilities.
&  \\
\hline

\multirow{3}{*}{\rotatebox[origin=c]{90}{\parbox{2.5cm}{\centering \textbf{G5. Interface}}}} 
& G5.1. Distribute disability features across the entire avatar interface rather than gathering them in a specialized category.
& G5.1. Distribute disability features across the entire avatar interface rather than gathering them in a specialized category. 
&  \\ \cline{2-4}

& G5.2. Use input controls that offer precise adjustments whenever possible. 
& G5.2. Use continuous controls for high-granularity customization. 
& Refined the wording to better describe PWD's customization control preferences. \\ \cline{2-4}

& G5.3. Offer an easy control to turn on/off or switch between disability features.
& G5.3. Offer an easy control to turn on/off or switch between disability features.
&  \\
\hline

\end{tabular}
\caption{The table compares the initial and revised guidelines to highlight changes. After revision, the initial set of 20 guidelines was refined and trimmed to 17 finalized design guidelines.}
\Description{}
\label{tab:changes}
\end{table*}

\afterpage{
\clearpage
\small

\begin{longtable}{p{0.4cm}p{4cm}p{6.5cm}p{5cm}}
\toprule
\textbf{} & \textbf{Design Guideline} & \textbf{Description} & \textbf{Examples and Demos} \\
\toprule
\endfirsthead

\toprule
\textbf{} & \textbf{Design Guideline} & \textbf{Description} & \textbf{Examples and Demos} \\
\toprule
\endhead


\multicolumn{4}{p{17cm}}{\change{\textbf{G0. Support disability representation in social VR avatars \cite{zhang2022, assets_24, chronic_pain_gualano_2024, kelly2023}.} Approximately 1.3 billion people experience significant disability, representing about 16\% of the global population \cite{WHO2023}. It's important to ensure PWD are included and represented equally in emerging technology such as social VR. As long as the platform involves avatar-based interactions, there is design space to support disability representation. The following set of guidelines can be flexibly applied to a variety of social VR platforms with different (1) avatar types (e.g., humanoid avatars in Rec Room \cite{recroom} vs. robotic-type avatars in Among Us \cite{amonguscharacters}), (2) aesthetic styles (e.g., life-like avatars in Horizon Worlds \cite{metaavatars} vs. abstract avatars in Roblox \cite{robloxwiki})), and (3) content focus (e.g., communication-heavy type in VRChat \cite{vrchat} vs. game-centric in Rec Room). We encourage practitioners to adopt G0 as a fundamental mindset when developing and designing avatars, considering it in the early stages, and consistently exploring opportunities to support disability representation.}}
\\ \midrule

\textbf{} & 
& \multicolumn{1}{c}{\textbf{G1. Avatar Body Appearance}} & 
\\ \midrule
\textbf{G1.1 \change{(HR)}}
& \textbf{Default to full-body avatars to enable diverse disability
representation across different body parts.}
& Avatar interfaces should offer full-body avatar options \cite{kelly2023}. \change{About 296,000 people in the U.S. live with paralysis of the lower half of the body, with around 17,900 new cases each year \cite{NSCISC2021}.} Given the \change{large affected user size and} dominant preferences for full-body avatars over others (e.g., upper-body only, or head and hands only), we recommend making it the default or the starting avatar template, giving users the maximum flexibility to further customize their avatars as they prefer. 
& \begin{minipage}[t]{\linewidth}
    E.g., A full-body avatar can show a prosthetic left foreleg. \\
    \includegraphics[width=3cm, height=2.8cm]{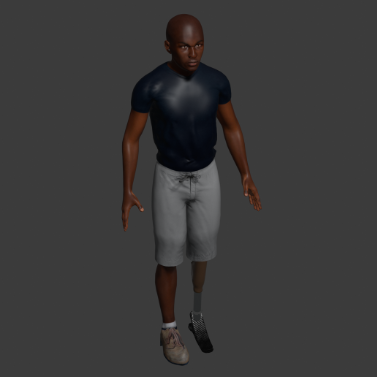}
  \end{minipage}
\\ \midrule

\textbf{G1.2 \change{(HR)}}
& \textbf{Enable flexible customization of body parts as opposed to using non-adjustable avatar templates.}
& Avatar interfaces should provide PWD sufficient flexibility to customize each avatar body part \cite{kelly2023}. \change{While the customization spans a wide range, the most commonly mentioned body parts to customize include (1) avatar height, (2) body shape, (3) limbs (i.e., number of limbs, length and strength of each limb), and (4) facial features (e.g., mouth shape, eye size).} Asymmetrical design options of body parts (e.g., eyes, ears) should also be available, such as changing size and direction of each eyeball to reflect disabilities like strabismus.
& \begin{minipage}[t]{\linewidth}
    E.g., Options that can customize the size of each eye. \\
    \includegraphics[width=3cm, height=2.7cm]{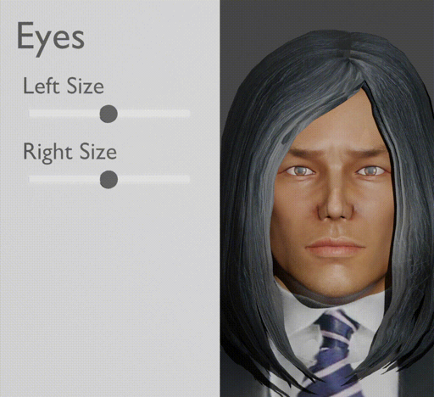}
  \end{minipage}
\\ \midrule

\textbf{G1.3 \change{(HR)}}
& \textbf{Prioritize human avatars to emphasize the ``humanity'' rather than the ``disability'' aspect of identity.}
& Social VR applications should offer human avatar options whenever the application theme allows.
& \begin{minipage}[t]{\linewidth}
    E.g., Human avatars that can show intersectional identities. \\
    \includegraphics[width=4cm, height=2cm]{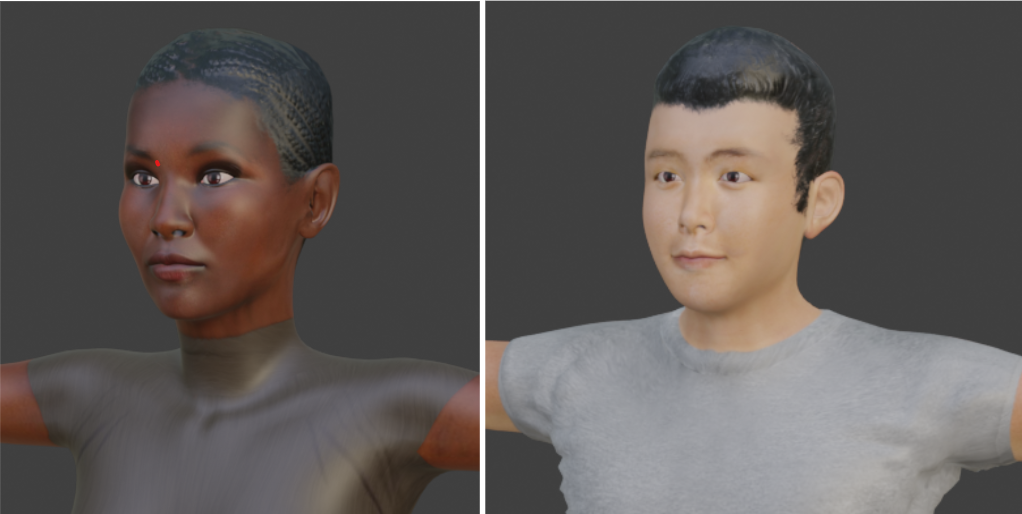}
  \end{minipage}
\\ \midrule

\textbf{G1.4 \change{(R)}}
& \textbf{Provide non-human avatar options to free users from social stigma in real life.}
& Besides human avatars, avatar interfaces should also provide diverse forms of non-human avatars, empowering PWD to choose the one they relate with flexibly.
& \begin{minipage}[t]{\linewidth}
    E.g., A robotic avatar representing left forearm amputation.\\
    \includegraphics[width=3cm, height=2.6cm]{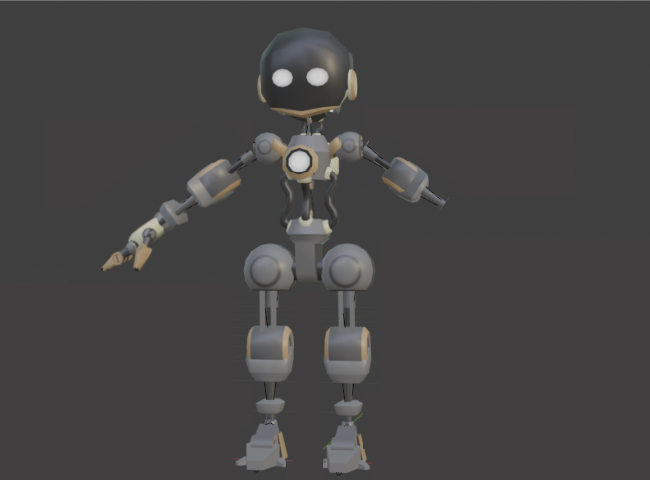}
  \end{minipage}
\\ \midrule

\textbf{} & 
& \multicolumn{1}{c}{\textbf{G2. Avatar Body Dynamics}} & 
\\ \midrule

\textbf{G2.1 \change{(HR)}}
& \textbf{Allow simulation or tracking of disability-related behaviors but only based on user preference.}
& Users should be able to control the extent of behavior tracking in social VR. With the advance of motion tracking techniques, avatar platforms may disable subtle behavior tracking by default to avoid disrespectful simulation, but allow users to easily adjust the tracking granularity for potential disability expression.
& \begin{minipage}[t]{\linewidth}
    E.g., Avatar can show motor tics (left) or not (right) based on the user's preference.\\
    \includegraphics[width=4cm, height=2cm]{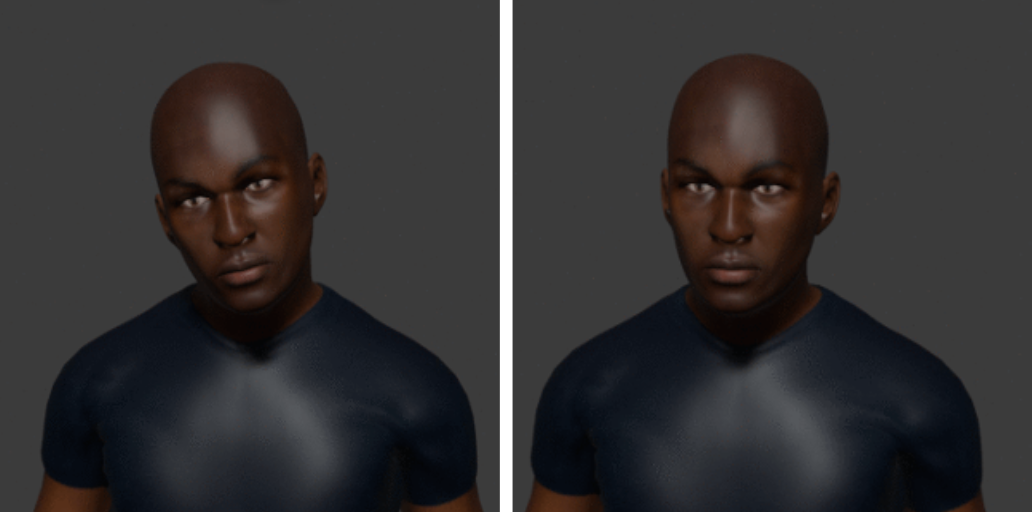}
  \end{minipage}
\\ \midrule

\textbf{G2.2 \change{(R)}}
& \textbf{Enable expressive facial animations to deliver invisible status.}
& Avatar platforms should enable diverse facial expressions, allowing PWD to express emotion, portray mental status, and indicate fluctuation of invisible disabilities \cite{assets_24}. \change{Including the five basic emotions is a good starting point, and practitioners should expand and diversify based on their own use scenarios. This guideline is particularly helpful for communication-heavy or mental therapy platforms, where expressing a range of emotions helps PWD convey their feelings more accurately.}
& \begin{minipage}[t]{\linewidth}
    E.g., Avatar shows diverse emotions, including anxiety (left), sadness (middle), and happiness (right).\\
    \includegraphics[width=4cm, height=1.9cm]{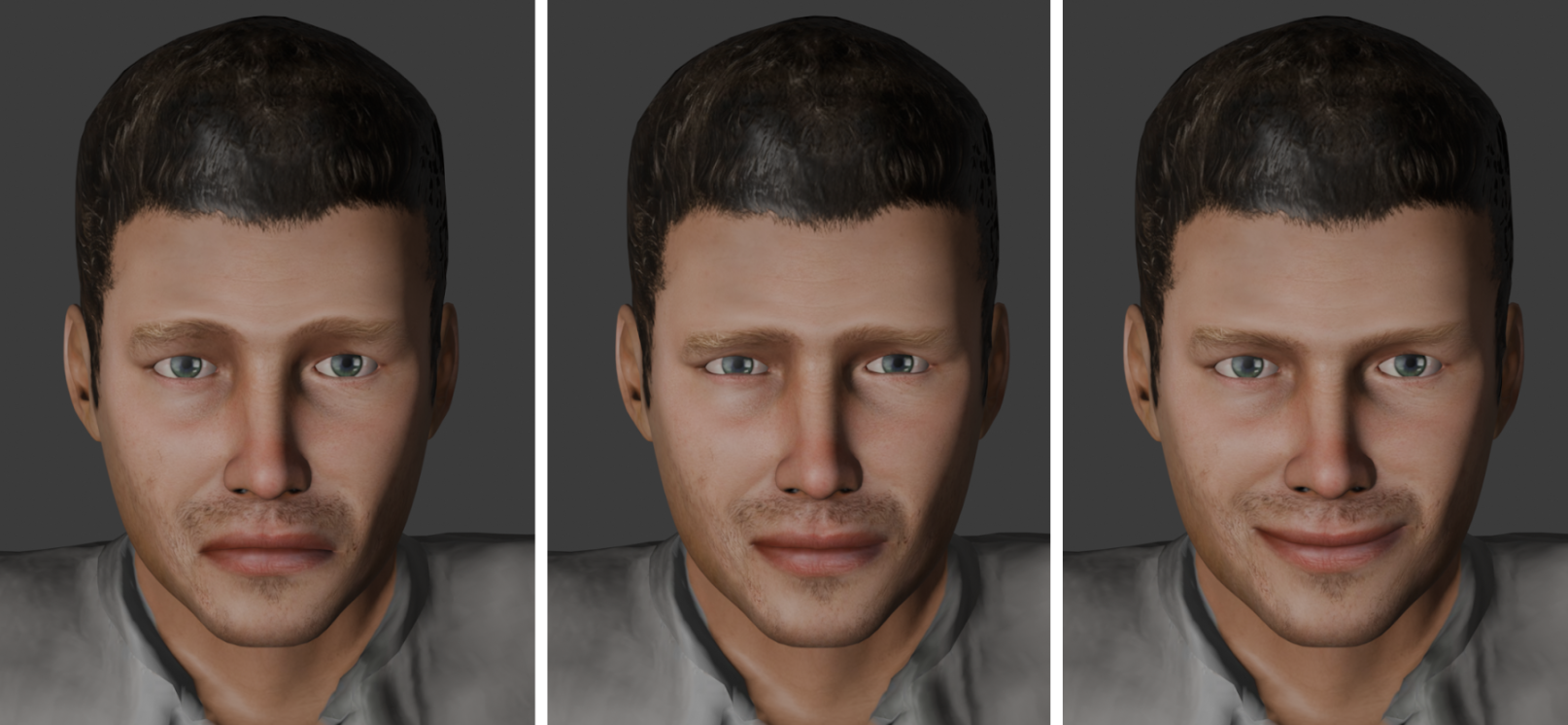}
  \end{minipage}
\\ \midrule

\textbf{G2.3 \change{(HR)}}
& \textbf{Prioritize equitable capability and performance over authentic simulation.}
& PWD value equitable and fair interaction experiences more than the authentic disability expression. Therefore, social VR platforms should ensure same level of capabilities and performance for all avatars no matter whether disability features or behaviors are involved. 
& \begin{minipage}[t]{\linewidth}
    E.g., The avatar with the wheelchair moves at the same speed as the avatar without the wheelchair.\\
    \includegraphics[width=3cm, height=2.2cm]{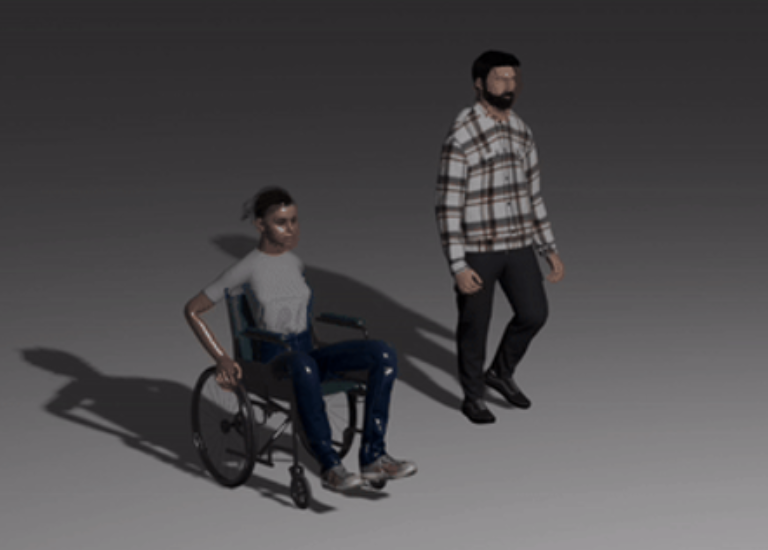}
  \end{minipage}
\\ \midrule

\textbf{} & 
& \multicolumn{1}{c}{\textbf{G3. Assistive Technology Design}} & 
\\ \midrule

\textbf{G3.1 \change{(HR)}}
& \textbf{Offer various types of assistive technology to cover a wide range of disabilities.}
& Avatar interfaces should offer assistive technologies that are commonly used by PWD \cite{zhang2022}. \change{Approximately 2.5 million people around the world are assistive technologies users \cite{LevelAccess2024}. According to our large-scale interview data, the most desired types of assistive technologies include: (1) mobility aids (e.g., wheelchair, cane, and crutches); (2) prosthetic limbs; (3) visual aids (e.g., white cane, glasses, and guide dog); (4) hearing aids and cochlear implants; and (5) health monitoring devices (e.g., insulin pumps, ventilator, smart watches). 
Practitioners should consider including at least these five categories of assistive technologies in avatar interfaces.
In addition, due to PWD's different technology preferences \cite{kelly_AI24}, we encourage practitioners to offer more than one assistive technology option within each category, for example, including guide dog, white cane, and glasses for visual aids.} 
& \begin{minipage}[t]{\linewidth}
    E.g., Offer multiple types of mobility aids, such as walking cane, crutches, and forearm crutches. \\
    \includegraphics[width=3cm, height=2.2cm]{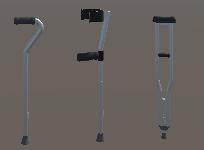}
  \end{minipage}
\\ \midrule

\textbf{G3.2 \change{(HR)}}
& \textbf{Allow detail customization of assistive technology for personalized disability representation.}
& Avatar platforms should allow customizations for assistive technology \change{\cite{kelly2023,zhang2022}. Basic customization options should include adjusting the colors of different assistive tech components and adding decorations (e.g., stickers, logos) to the assistive technologies. More customization could be added based on specific use cases.} 
& \begin{minipage}[t]{\linewidth}
    E.g., Users can adjust the color of the power wheelchair, like the cushions, wheels, and chassis cover.\\
    \includegraphics[width=4cm, height=2cm]{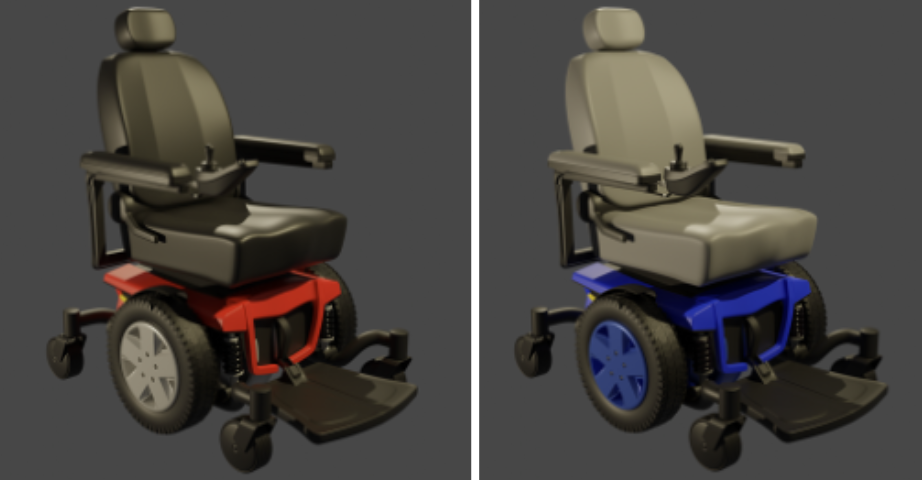}
  \end{minipage}
\\ \midrule

\textbf{G3.3 (HR)}
& \textbf{Provide high-quality, authentic simulation of assistive technology to present disability respectfully and avoid misuse.}
& To avoid misunderstandings or misuse, the assistive tech simulation should convey standardized, \change{authentic details of the real-world assistive devices \cite{zhang2023}, regardless the overall avatar style. For example, the design of a white cane should show the details of tip and follow its standardized color selection, no matter the design style is photorealistic or cartoon. We recommend practitioners to model assistive technologies by following their established design standards, such as design guidelines for white canes \cite{who_white_canes}, wheelchairs \cite{russotti_ansi_wheelchairs}, and hearing devices \cite{ecfr_800_30}.}
& \begin{minipage}[t]{\linewidth}
    E.g., Thh wheelchair should be designed with high-fidelity and realistic details to represent disability respectfully.\\
    \includegraphics[width=3cm, height=2.4cm]{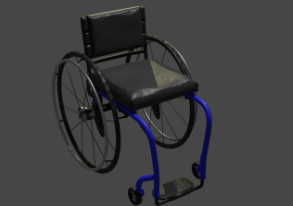}
  \end{minipage}
\\ \midrule

\textbf{G3.4 \change{(R)}}
& \textbf{Demonstrate the liveliness of PWD through dynamic interactions with assistive technology.}
& Beyond providing assistive tech options, social VR platforms should enable suitable interactions between avatars and assistive tech. The interactions should authentically reflect PWD's real-world usage of their assistive tech, such as how a blind user sweeps their cane, or how a wheelchair user moves their arms to control their wheelchair.
& \begin{minipage}[t]{\linewidth}
    E.g., Avatar controls the manual wheelchair through pushing. \\
    \includegraphics[width=3cm, height=2.8cm]{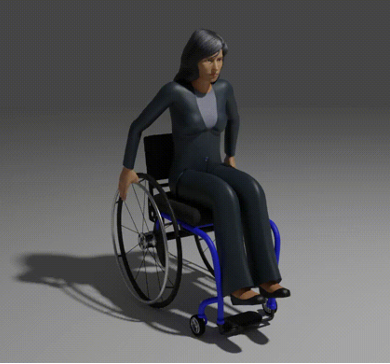}
  \end{minipage}
\\ \midrule

\textbf{G3.5 \change{(HR)}}
& \textbf{Avoid overshadowing the avatar body with assistive technology.}
& The size of assistive technology should not dominate the avatar body but rather fit the body size. \change{Avatar platforms should automatically match the assistive tech model to different avatar body sizes by default, and allow users to adjust the size of assistive technology to achieve their preferred avatar-aid ratio. Moreover, the combination of avatar and assistive tech should be seamless without affecting the quality and aesthetics of the original avatar \cite{kelly2023}.}
& \begin{minipage}[t]{\linewidth}
    E.g., Users can change the size of assistive technology to match with their avatar. \\
    \includegraphics[width=3cm, height=3cm]{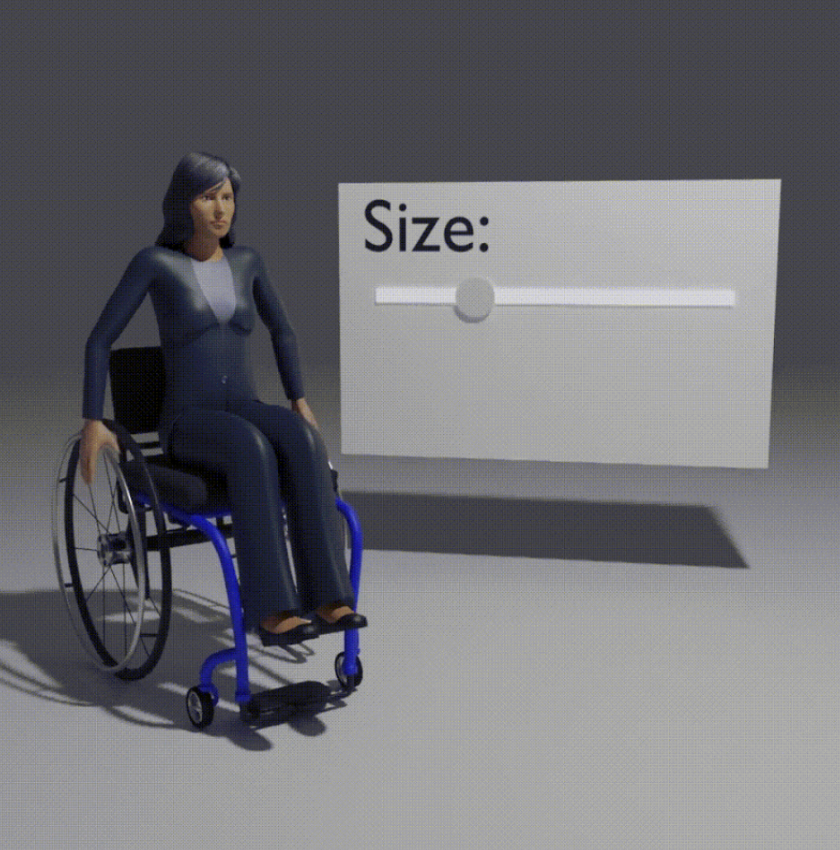}
  \end{minipage}
\\ \midrule

\textbf{} & 
& \multicolumn{1}{c}{\textbf{G4. Peripherals around Avatars}} & 
\\ \midrule

\textbf{G4.1 \change{(HR)}}
& \textbf{Provide suitable icons, logos, and slogans that represent disability communities.}
& Awareness-building items or presets (e.g., logos, slogans) should be provided, allowing users to attach them to various areas on or around the avatars, such as the apparel, accessories, and assistive tech \cite{assets_24, zhang2022}. \change{We compiled a list of widely recognized and preferred symbols that represent different disabilities for practitioners to refer to, including (1) the rainbow infinity symbol the represents the autism community \cite{assets_24, rainbow_infinity_symbol}, (2) the sunflower that represents hidden disabilities \cite{isit_assets24, hidden_disability_sunflower}, (3) the disability pride flag \cite{disability_pride_flag}, (4) the spoons, symbolizing spoon theory for people with chronic illness \cite{kelly2023, assets_24}, and (5) the zebra symbols for rare diseases \cite{assets_24, Gualano_2023}.} 
& \begin{minipage}[t]{\linewidth}
    E.g., An avatar wearing a T-shirt with a rainbow and infinity symbol to represent the autism spectrum disorder community. \\
    \includegraphics[width=3cm, height=2.8cm]{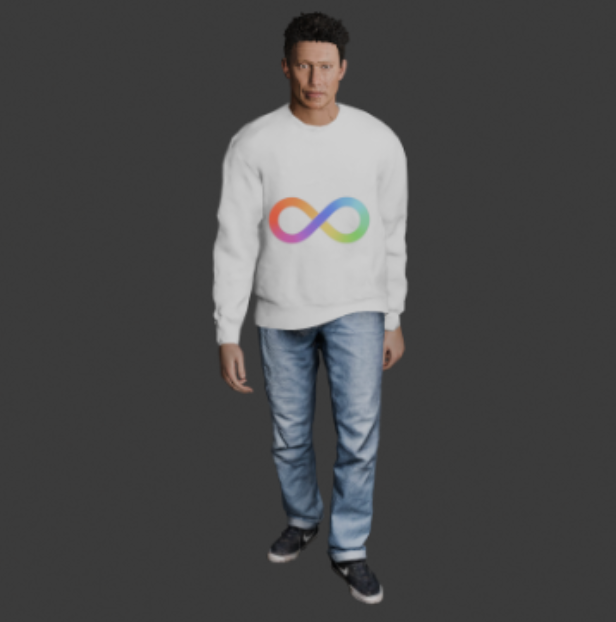}
  \end{minipage}
\\ \midrule

\textbf{G4.2 \change{(R)}}
& \textbf{Leverage spaces beyond the avatar body to present disabilities.}
& When designing avatars, developers and designers should consider leveraging avatar's peripheral space to enable users to better express their status, especially for individuals with invisible disabilities. \change{Some design examples include a weather background to indicate mood and a battery sign to indicate energy level \cite{assets_24}.} 
& \begin{minipage}[t]{\linewidth}
    E.g., An avatar with a floating bubble overhead, showing a level of social energy. \\
    \includegraphics[width=4cm, height=2.4cm]{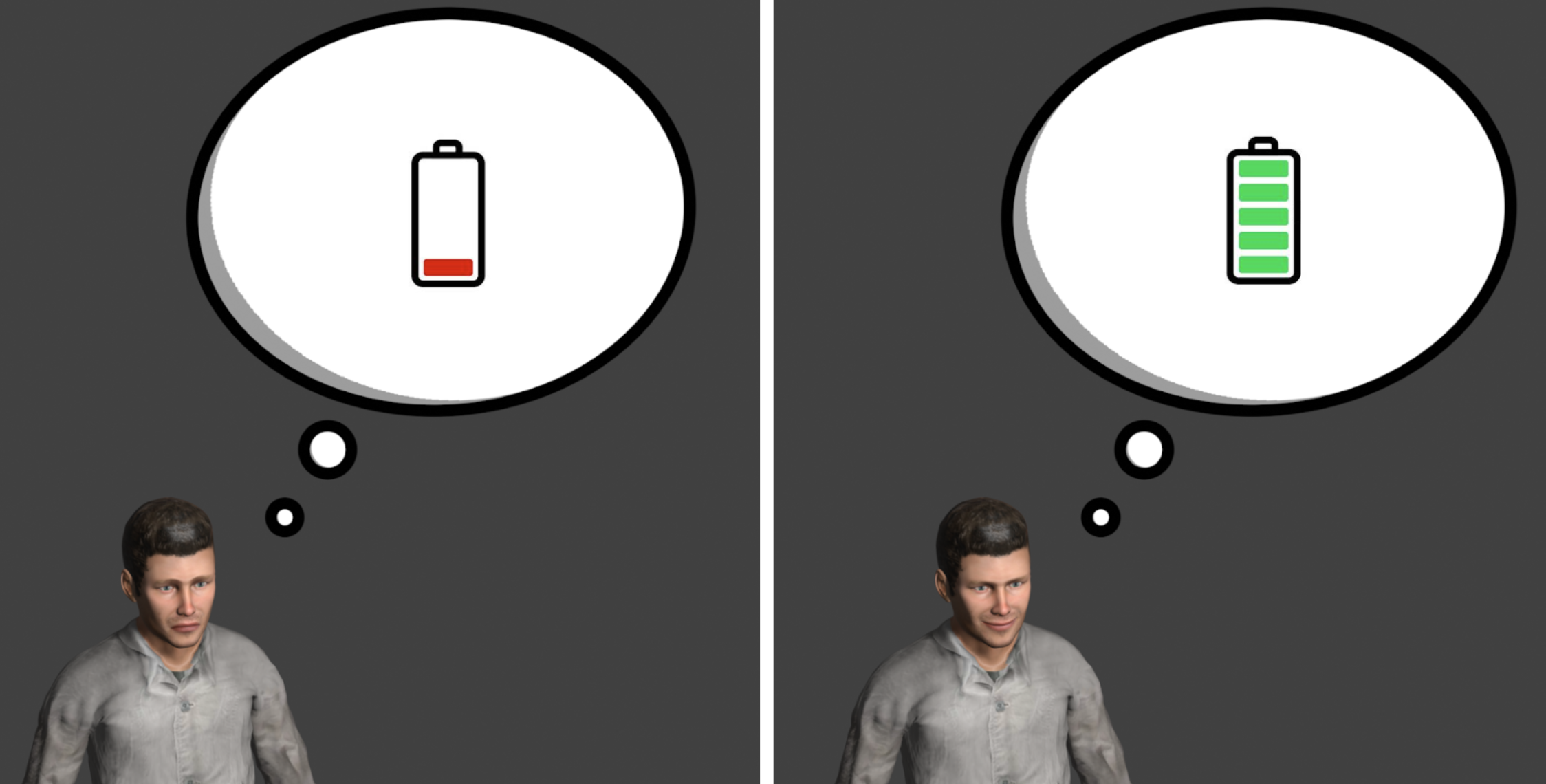}
  \end{minipage}
\\ \midrule

\textbf{} & 
& \multicolumn{1}{c}{\textbf{G5. Interface and Control}} & 
\\ \midrule

\textbf{G5.1 \change{(HR)}}
& \textbf{Distribute disability features across the entire avatar interface rather than gathering them in a specialized category.}
& Avatar features for disability expression should be treated in the same way as other avatar features. In avatar interfaces, disability-related features should be properly distributed in their corresponding categories. \change{There should not be a specialized category for PWD. For example, assistive technologies should be included in the accessory category rather than an assistive tech category.} 
& \begin{minipage}[t]{\linewidth}
    E.g., Walking canes and wheelchairs are included under the accessory category along with items like glasses, hats and bags. \\
    \includegraphics[width=3cm, height=3cm]{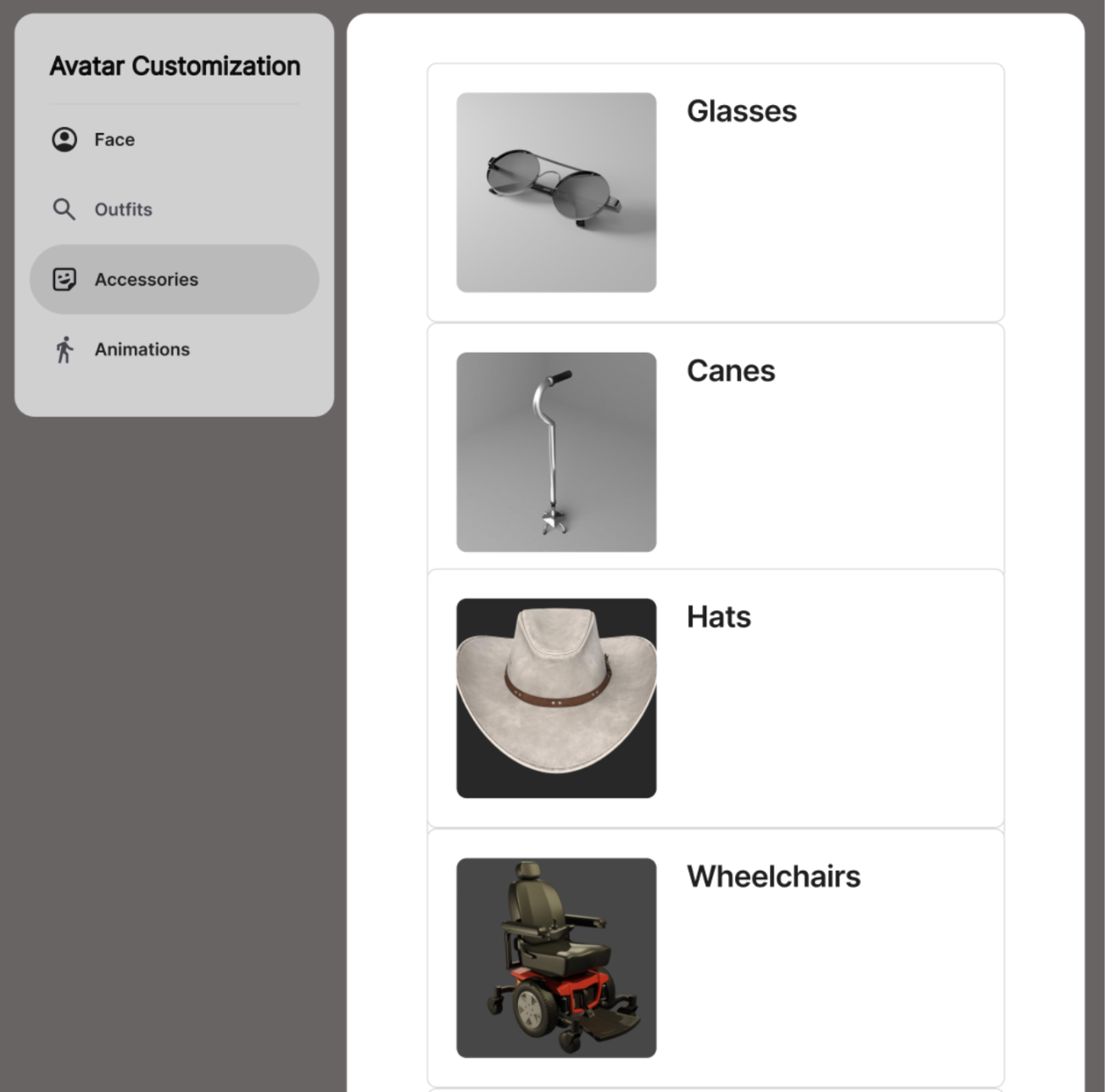}
  \end{minipage}
\\ \midrule

\textbf{G5.2 \change{(HR)}}
& \textbf{Use continuous controls for high-granularity customization.}
& Avatar interfaces should adopt input controls that offer a continuous range of options to enable flexible customization. \change{This could be widely applied to a variety of design attributes, such as the size and shape of multiple avatar body parts.}
& \begin{minipage}[t]{\linewidth}
    E.g., Offer a slider to change the length of limbs. \\
    \includegraphics[width=3cm, height=2.7cm]{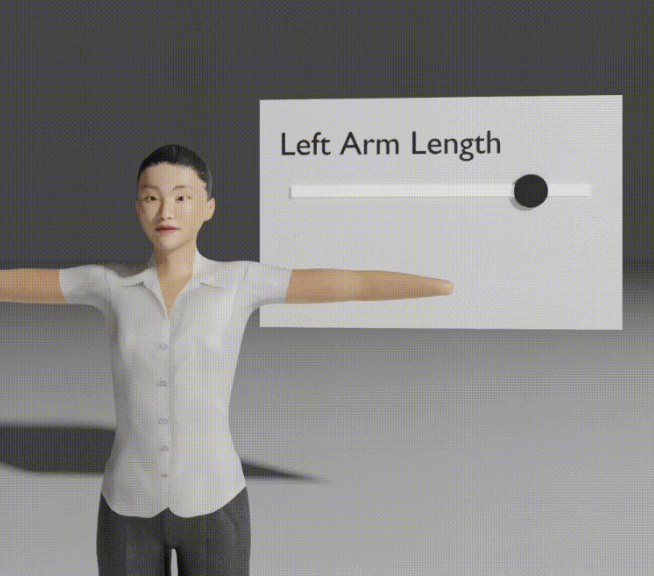}
  \end{minipage}
\\ \midrule

\textbf{G5.3 \change{(HR)}}
& \textbf{Offer an easy control to turn on/off or switch between disability features.}
& Social VR platforms should provide easy-to-access shortcut control enable users to conduct \change{\textit{ad-hoc} avatar updates on the go. Important control functions include: (1) toggling on and off the disability-related features \cite{assets_24}; (2) switching between different saved avatars \cite{kelly2023}; and (3) updating status for fluctuating conditions \cite{assets_24}.} 
& \begin{minipage}[t]{\linewidth}
    E.g., Users should be able to turn disability-related features on and off with a single click. \\
    \includegraphics[width=4cm, height=1.5cm]{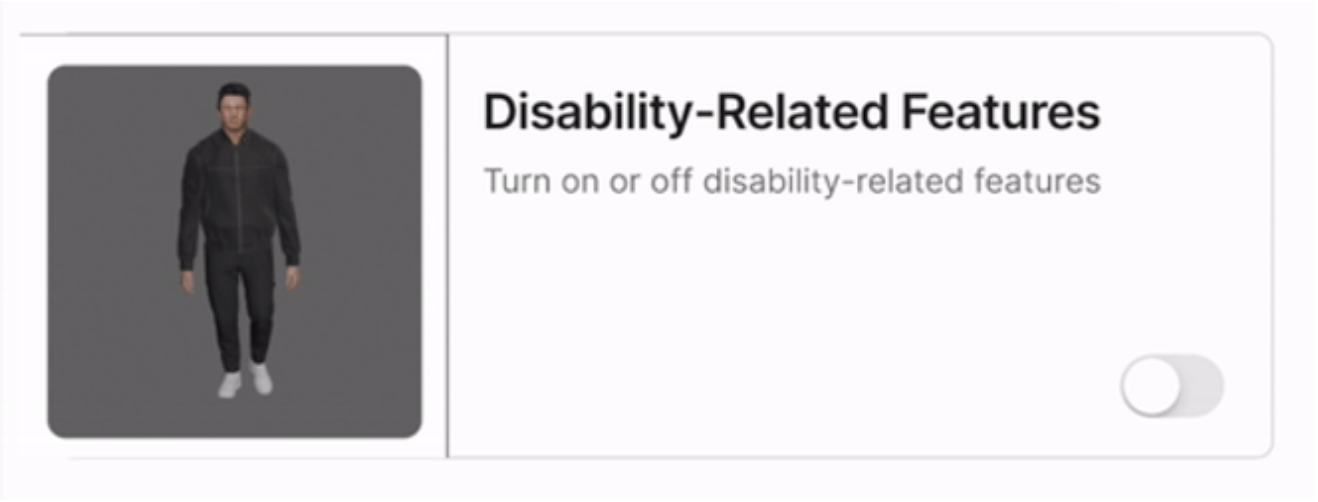}
  \end{minipage}
\\  
\bottomrule
\caption{\change{A full version of our \textit{revised} guidelines that incorporated VR experts' feedback. We present a set of 17 inclusive avatar guidelines for PWD that covered five avatar design aspects. Each guideline includes: the guideline statement with a recommendation level (Highly Recommended (HR) or Recommended (R)), a detailed description of guideline, and an actionable implementation example.}}
\Description{This table, titled "Table 7," provides a detailed overview of 17 revised guidelines for inclusive avatar design. It includes three columns: "Design Guideline," "Description," and "Examples and Demos," organized into five specific guideline sections. Each guideline has a recommendation level denoted as ``HR'' for Highly Recommended or ``R'' for Recommended.}
\label{tab:full_revised}
\end{longtable}
}


\end{document}